\documentclass[journal]{IEEEtran}
\usepackage{amsmath,amsfonts}
\usepackage{algorithmic}
\usepackage{array}
\usepackage[caption=false,font=footnotesize,labelfont=rm,textfont=rm]{subfig}

\usepackage{multirow}
\usepackage{textcomp}
\usepackage{stfloats}
\usepackage{graphicx}
\usepackage{url}
\usepackage{verbatim}
\usepackage{cite}
\usepackage{makecell}
\usepackage{array}
\usepackage{multirow}
\usepackage{graphbox}
\usepackage{endnotes}
\usepackage{color}
\usepackage[table]{xcolor}
\usepackage{array}
\usepackage{xcolor}
\usepackage{amsthm}
\usepackage{tabularx}
\newtheorem{theorem}{Theorem}

\theoremstyle{remark}
\def\BibTeX{{\rm B\kern-.05em{\sc i\kern-.025em b}\kern-.08em
    T\kern-.1667em\lower.7ex\hbox{E}\kern-.125emX}}
\usepackage{balance}
\begin{document}
\newcolumntype{L}[1]{>{\raggedright\arraybackslash}p{#1}}
\newcolumntype{C}[1]{>{\centering\arraybackslash}p{#1}}
\newcolumntype{R}[1]{>{\raggedleft\arraybackslash}p{#1}}

\title{Explainable Gated Bayesian Recurrent Neural Network for Non-Markov State Estimation}
\author{Shi Yan, Yan Liang$\dagger$, Le Zheng,~\IEEEmembership{Senior Member,~IEEE}, Mingyang Fan, Xiaoxu Wang,~\IEEEmembership{Member,~IEEE}, \\ Binglu Wang,~\IEEEmembership{Member,~IEEE}
		% <-this % stops a space
         \thanks{Shi Yan, Yan Liang, Mingyang Fan, Xiaoxu Wang and Binglu Wang are with the School of Automation, Northwestern Polytechnical University, Xi'an 710072, China (e-mail: yanshi@mail.nwpu.edu.cn; liangyan@nwpu.edu.cn; van\_my@163.com; wbl921129@gmail.com; woyaofly1982@nwpu.edu.cn).}
        \thanks{Le Zheng is with the Radar Research Laboratory, School of Information and Electronics, Beijing Institute of Technology, Beijing 100081, China (e-mail: le.zheng.cn@gmail.com).}
        \thanks{$\dagger$Corresponding author: Yan~Liang.}
	\thanks{This work was supported by the National Natural Science Foundation of China under Grant 61873205.}}

\markboth{Journal of \LaTeX\ Class Files,~Vol.~18, No.~9, September~2020}%
{How to Use the IEEEtran \LaTeX \ Templates}

\maketitle
\begin{abstract}
The optimality of Bayesian filtering relies on the completeness of prior models, while deep learning holds a distinct advantage in learning models from offline data. Nevertheless, the current fusion of these two methodologies remains largely ad hoc, lacking a theoretical foundation. This paper presents a novel solution, namely an explainable gated Bayesian recurrent neural network specifically designed to state estimation under model mismatches. Firstly, we transform the non-Markov state-space model into an equivalent first-order Markov model with memory. It is a generalized transformation that overcomes the limitations of the first-order Markov property and enables recursive filtering. Secondly, by deriving a data-assisted joint state-memory-mismatch Bayesian filtering, we design a Bayesian gated framework that includes a memory update gate for capturing the temporal regularities in state evolution, a state prediction gate with the evolution mismatch compensation, and a state update gate with the observation mismatch compensation. The Gaussian approximation implementation of the filtering process within the gated framework is derived, taking into account the computational efficiency. Finally, the corresponding internal neural network structures and end-to-end training methods are designed. The Bayesian filtering theory enhances the interpretability of the proposed gated network, enabling the effective integration of offline data and prior models within functionally explicit gated units. In comprehensive experiments, including simulations and real-world datasets, the proposed gated network demonstrates superior estimation performance compared to benchmark filters and state-of-the-art deep learning filtering methods.
\end{abstract}
\section{Introduction}\IEEEPARstart{B}{ayesian} filtering (BF) constructs a posterior probability density function (PDF) for the system state by leveraging available information \cite{li2023finite}, which is widely used in target tracking \cite{Haoxiaohui,Tracking1,Trackingxu}, navigation \cite{navigation_intro,hao_navigation}, and localization \cite{localization_intro} due to its solid theoretical foundation and effective computational design. During the filtering process, state-space models (SSMs) play a pivotal role in depicting state evolution and sensor observations. The optimality of traditional Bayesian filters depends on the correctness of prior SSMs \cite{KF_intro}, and hence their performance significantly deteriorates in the presence of model mismatches, while the complete and correct prior SSM is difficult to obtain in a complex and non-cooperative environment. Therefore, much attention has been paid to online Bayesian optimization, with representative approaches being the multi-model (MM) estimation and joint estimation identification (JEI). Its idea is the real-time inference of model parameters along with the recursive computation of density. The limited information available in online measurements necessitates certain prior conditions or assumptions for these parameters. Specifically, the MM estimation requires a prior model set and transition probability matrix \cite{MM_overview,IMM,HGMM}; JEI implemented by the expectation-maximization (EM) for parameter identification assumes that the estimated parameter is piecewise constant \cite{EMTSP,lan2016survey,liu2021based,liu2022maneuvering}; and JEI implemented by the variational inference (VI) for covariance identification \cite{VBTSP,VBI_huang,zhang2021variational,VB2022} assume that the estimated covariance has a conjugate distribution. More importantly, the online Bayesian optimization approaches exclusively deal with real-time measurement data, while disregarding the abundant offline data.

With the widespread deployment of sensors and advancements in simulation technologies, there has been a notable increase in the available offline data, leading to a surge in the use of deep learning (DL). By capturing the temporal regularities in state evolution through their powerful nonlinear fitting capabilities and memory iteration mechanisms, various gated recurrent neural networks (RNNs), like long short-term memory (LSTM) \cite{LSTM_ori} and gated recurrent unit (GRU) \cite{GRU_ori}, have demonstrated remarkable performance in handling time series data \cite{lstm1,lstm2}. Gated RNNs are utilized to establish mapping functions from measurements to states, revealing the feasibility of extracting implicit modeling information from offline data \cite{pure_lstm1,pure_lstm2}. Nonetheless, pure DL approaches do not systematically integrate prior model knowledge, which necessitates a large number of learnable parameters and a substantial amount of labeled data, often resulting in a lack of interpretability.

To introduce prior model knowledge and improve interpretability, methods combining DL and BF have been widely studied in recent years. The RNN is utilized for dynamic model learning, i.e., it is integrated into the Bayesian filter as a mapping for the state transition \cite{GRU_pf,CNN_pf,jung2020mnemonic}, while this leads to the neglect of the prior dynamic model knowledge. This idea of utilizing RNN to directly learn the dynamic and measurement models of the Bayesian filter has more applications in the computer vision field \cite{ICCV2017,TNNLS_Dynanet}. An improved approach employs the LSTM to correct the estimated state of a model-based (MB) Bayesian filter \cite{DeepMTT}. When the model is severely mismatched, such an ad hoc combination faces failure as the filter diverges, and the covariance is not corrected so that it is inconsistent with the corrected state. Given the significance of filter gain in balancing prediction and measurement, a recent study utilized the GRU to estimate the gain \cite{KalmanNet}. By omitting second-order moment computations, this method accelerates the computation but sacrifices the utilization of prior noise statistical information and the ability to output covariance. In general, it is essential to derive an algorithmic framework based on a well-defined model and Bayesian filtering theory when introducing DL in the filtering. This is a crucial element currently lacking in existing methods, and it has the potential to enhance interpretability and estimation performance.

%From an algorithmic architecture perspective, rather than the complex process of embedding gated RNNs into the filter, a more efficient strategy is to construct an underlying gated RNN tailored to state estimation. The design of internal gated structures for gated RNNs tailored to different task characteristics has been widely studied in recent years, and such designs simplify the computational process while reducing the learnable parameter scales \cite{UGRNN,FAST_GRNN,Oth2}. While the task specific gated RNNs have been successfully applied in fields such as language processing \cite{AAAI_GRNN} and image processing \cite{TGRS_GRNN,Att_GRNN_TIP,Att_GRNN_SPL}, a gated RNN specifically designed for state estimation has yet to be explored. In addition, existing gated RNNs generally lack interpretability, as their inner mechanisms remain insufficiently understood \cite{hou2020learning}. This paper introduces a novel perspective to the field by proposing a well-interpretable gated RNN customized for state estimation.

To this end, we propose a general state estimation problem for the non-Markov system, considering the non-Markov state evolution model, uncertain measurement model, and labeled offline data. For this problem, an underlying explainable gated Bayesian RNN (EGBRNN) is designed. At first, it is discovered that the non-Markov SSM can be transformed into an equivalent first-order Markov model with memory through function nesting, which enables efficient recursive computation while overcoming the limitations of the first-order Markov property. Secondly, a data-assisted joint state-memory-mismatch BF framework is derived based on the transformed model, featuring three well-defined gated units: the memory update gate captures the temporal regularities in state evolution, as well as the state prediction and state update gates perform filtering while compensating for model mismatches. With its functionally explicit gated units, the framework efficiently integrates offline data and prior model knowledge with good interpretability. Considering computing efficiency, the framework is derived as a Gaussian approximation implementation. Finally, the internal neural network (NN) structure and end-to-end training method of each gated unit are designed. Notably, rather than artificially combing the DL and the Bayesian filter, the proposed EGBRNN is naturally derived from BF theory. In comprehensive experiments with benchmark simulations and real-world datasets, the EGBRNN outperforms various MB filters and state-of-the-art (SOTA) filtering methods.

Notation: throughout this paper, the superscript ${\left(  \cdot  \right)^\top}$ denotes the transpose operation and the superscript ${\left(  \cdot  \right)^{-1}}$ denotes the inverse operation; ${\left(  \cdot  \right)}$ represents identical content to that stated in the preceding parenthesis; ${\rm{E}}\left[ { \cdot | \cdot } \right]$ denotes the conditional expectation; $\left \| \cdot \right \| ^{2}$ represents the two-norm operation;  ${\rm{diag}}\left(a, b\right)$ represents a diagonal matrix with diagonal elements $a$ and $b$.

\section{Problem Formulation}
The state estimation of dynamic systems typically involves the utilization of the following first-order Markov SSM with additive noises.
\begin{flalign}
&\noindent \emph{System model:}&\notag \\
&\quad \quad \quad \quad \quad \quad \quad  {{\bf{x}}_k} = f_k\left( {{{\bf{x}}_{k - 1}}} \right) + {\bf{w}}_{k}&
\label{system_model}\\
&\noindent \emph{Measurement model:}&\notag \\
&\quad \quad \quad \quad \quad \quad \quad  {{\mathbf{z}}_k} = {h_k}\left({{\mathbf{x}}_k} \right) + {{\mathbf{v}}_k}&
\label{measurement_model}
\end{flalign}
where the subscript $k$ represents the time index; ${\bf x}_k$ is the state vector, such as the aircraft's position and velocity; ${\bf z}_k$ is the measurement vector, such as the radial distance and azimuth of radar observation; $f_k$ and $h_k$ are the nominal linear/nonlinear state-evolution and state-measurement functions, respectively; the process noise ${\bf{w}}_{k}$ and the measurement noise ${{\bf{v}}_k}$ are independent of each other, and their PDFs are known and always being Gaussian distribution.

The first-order Markov property of the system model in Eq. (\ref{system_model}) has certain reasonable aspects. At first, the most recent state has the most effect on the evolution since the state evolution could be roughly described as an accumulative process. Secondly, the state transition process between adjacent frames is easy to analyze, making it possible to characterize the process with a first-order model. Finally, the first-order Markov property supports the recursive computation. The above advantages have led to the widespread application of this modeling form.

\begin{figure}[t]
\centering
\includegraphics[width=1.0\linewidth]{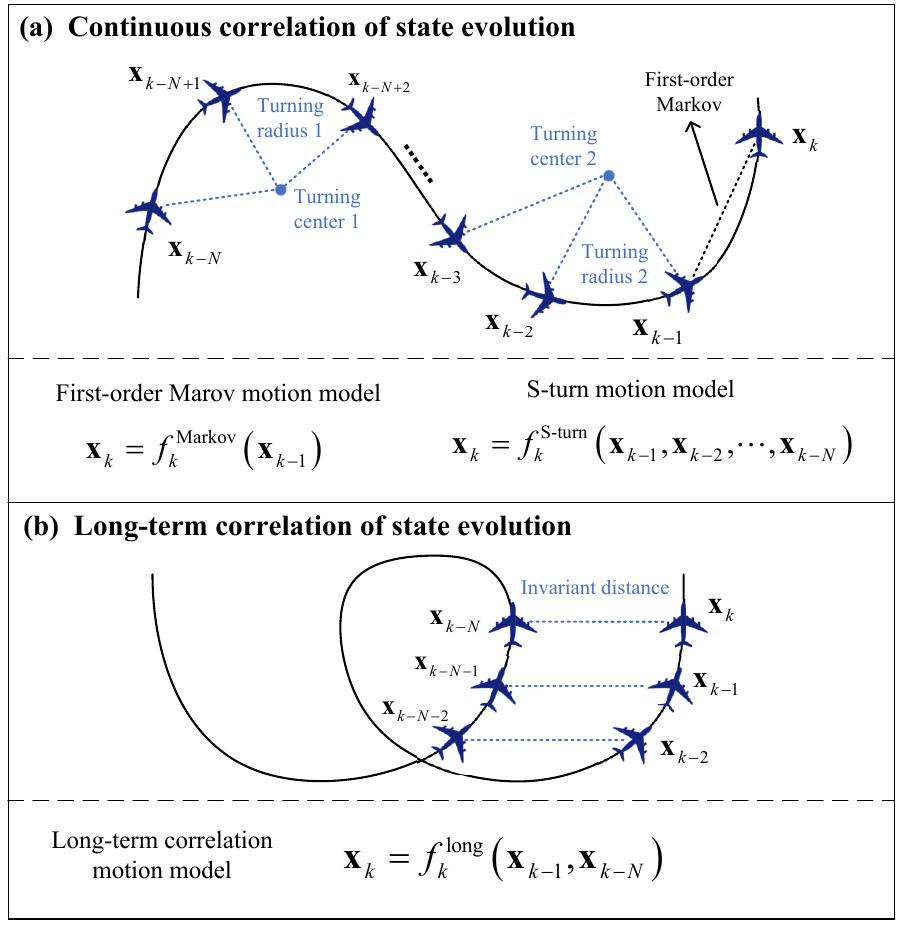}
\caption{{\textbf{Examples of the non-Markov state evolution model.} (a) The S-turn motion model requires multiple frames of state to be recognized. (b) The current state can be obtained by analyzing historical states.}}
\label{introducion_pic}
\end{figure}

In fact, the real-world state evolution may be more complex than the model in Eq. (\ref{system_model}). In practical state estimation tasks such as target tracking and navigation, the state evolution could be related to any of the historical states. 
From the aspects of continuous correlation and long-term correlation of state evolution, Fig. \ref{introducion_pic} shows two representative types of non-Markov models in target tracking.
In Fig. \ref{introducion_pic} (a), accurately modeling the target's S-turn motion requires identifying two turning centers and two turning radii.  
These parameters can be computed by multiple previous states to construct an accurate S-turn motion model $f^{{\rm{S - turn}}}_k$, while the exact number of states needed varies across different frames, dependent on the maneuver's unknown start time, $k-N$. 
The first-order Markov model motion $f^{\rm{Markov}}_k$ is clearly inaccurate, being unable to compute the parameters of such a complex motion. Moreover, the target's motion involves a combination of recurring typical motion modes, allowing for the determination of the current state based on historical states due to the invariance caused by the same motion mode. As shown in Fig. \ref{introducion_pic} (b), the invariant distance allows the long-term correlation motion model $f^{\rm{long}}_k$ to obtain ${\bf x}_k$ by ${\bf x}_{k-1}$ and ${\bf x}_{k-N}$, but $N$ is unknown. Such a non-Markov property is also prevalent in natural language processing. For instance, in a sentence, the meaning of "bank" depends on the preceding context, with interpretations varying between a financial institution and the edge of a river or lake. Obviously, the state evolution could not be accurately depicted by the first-order Markov model, and it should be modeled in a non-Markov form \cite{nonMarkov}, which is described as
\begin{align}
{{\bf{x}}_k} = f_k^{\text{general}}\left( {{{\bf{x}}_{k - 1}},{{\bf{x}}_{k - 2}}, \cdots ,{{\bf{x}}_1}} \right)+{\bf w}_k
\label{general_model}
\end{align}
where the general state-evolution function $f_k^{\text{general}}$ reflects the effects of previous states and cannot be given a priori due to the increasing dimensionality. Note that Eq. (\ref{system_model}) is a particular case of Eq. (\ref{general_model}) when the state sequence in $f_k^{\text{general}}$ contains only the previous single frame.

The measurement model in Eq. (\ref{measurement_model}) is also uncertain in practical state estimation tasks, and the uncertainties mainly come from two aspects: 1) the model mismatch caused by observation anomalies, such as offsets or unknown rotations in the state observation function due to poor sensor calibration; 2) it is often necessary to approximate intractable complex mappings with computationally convenient functions, which causes errors, e.g., the linearization of the non-linear system. We define the actual measurement model as
\begin{align}
{{\mathbf{z}}_k} = {h_k^\text{general}}\left({{\mathbf{x}}_k} \right) + {{\mathbf{v}}_k}
\label{general_meas}
\end{align}
where the general state-measurement function $h_k^{\text{general}}$ is unknown, which depicts the actual observation process.

For the filtering of the uncertain non-Markov SSMs presented in (\ref{general_model}) and (\ref{general_meas}), we also consider access to labeled offline data. The abundant regularity information about state evolution and sensing observations is implicit in the offline data, which needs to be mined using time-series DL methods. Here, the available offline data consists of some sequence of observations and their corresponding ground truth states, i.e.,
\begin{flalign}
{\cal D} = {\left\{ {{\bf{x}}_{1:K}^i,{\bf{z}}_{1:K}^i} \right\}_{i = 1 \cdots I}}&
\label{offline_data}
\end{flalign}
where ${\bf{x}}_{1:K}^i$ is the state ground-truth sequence obtained offline and ${\bf{z}}_{1:K}^i$ is the corresponding measurement sequence; $K$ is the duration of the system evolution; and $I$ is the number of samples in the data set. Such an offline data set can be accumulated by collecting sensor readings or executing simulations. For instance, collecting location ground-truth and odometer readings for wheeled robots.

It is quite challenging for the filtering to SSMs in Eqs. (\ref{general_model}) and (\ref{general_meas}) with considering the offline data set ${\cal D}$. On the one hand, computational cost considerations require the derivation of efficient recursive estimation frameworks for non-Markov SSMs. On the other hand, it is required to extract regularity information about state evolution and sensing observations implicit in the offline data, which is used to learn uncertain SSMs. In addition, one needs to consider the prior information that can be available while utilizing the offline data, i.e., the prior nominal model in Eqs. (\ref{system_model}) and (\ref{measurement_model}). To this end, our goal is to construct a state estimation algorithm for uncertain non-Markov SSMs by synthesizing prior model information and offline data. This algorithm should have an efficient recursive estimation framework derived from BF theory.

\section{Explainable Gated Bayesian Recurrent Neural Network}
The algorithm's design comprises the following three stages. At first, the non-Markov SSM is transformed into an equivalent first-order Markov model with memory. Secondly, a data-assisted joint state-memory-mismatch BF is derived based on the transformed model, leading to an explainable gated framework. The filtering process within the gated framework is implemented using the Gaussian approximation. Finally, we design the structure of the NNs in each gated unit and the corresponding training method.

\subsection{Problem transformation}
\label{Method_A}
By introducing prior nominal models $f_k$ and $h_k$ in Eqs. (\ref{system_model}) and (\ref{measurement_model}), the problem of state estimation under uncertain and non-Markov models is transformed into the processing of the corresponding errors, thus we have
\begin{align}
{{\mathbf{x}}_k} =& f_k\left( {{{\mathbf{x}}_{k - 1}}} \right) + \underbrace {f_k^{{\text{general}}}\left( {{{\mathbf{x}}_{k - 1}}, \cdots ,{{\mathbf{x}}_1}} \right) - f_k\left( {{{\mathbf{x}}_{k - 1}}} \right)}_{\Delta _k^f}
+ {{\mathbf{w}}_k}
\label{sys_error}\\
{{\mathbf{z}}_k} =& h_k\left( {{{\mathbf{x}}_k}} \right) + \underbrace {{h_k^{{\text{general}}}}\left( {{{\mathbf{x}}_k}} \right) - h_k\left( {{{\mathbf{x}}_k}} \right)}_{\Delta _k^h} + {{\mathbf{v}}_k}
\label{meas_error}
\end{align}
where $\Delta _k^f$ and $\Delta _k^h$ are the corresponding state-dependent evolution mismatch error and observation mismatch error, respectively.

In Eq. (\ref{sys_error}), the non-Markov error ${\Delta _k^f}$ is difficult to recognize directly since it continues to change as the states accumulate. Here, a two-layer nested function is constructed to approximate ${\Delta _k^f}$, which is described as
\begin{align}
\Delta _k^f = f^{\Delta}\left( {\underbrace {g_k^{\Delta}\left( {{{\mathbf{x}}_1}, \cdots ,{{\mathbf{x}}_{k - 1}}} \right)}_{{{\mathbf{c}}_k}}} \right)
\label{delta_f}
\end{align}
where $g_k^{\Delta}$ encapsulates the effects of historical states, representing a regular memory pattern in state evolution, and hence denoted as memory ${\bf c}_k$, which exhibits non-Markov characteristics; and $f^{\Delta}$ is an output function. In complex state evolution, ${\bf c}_k$ is essentially a hidden variable that is difficult to construct explicitly into a physical meaning, while the extraction regularity of the memory is implicit in a large amount of offline data. 

For the memory ${\bf c}_k$, considering the evolutionary regularities implicit in the states at different instants are similar, we also use the idea of nested functions to deal with it. Specifically, ${\bf c}_k$ is approximated through the following nested function:
\begin{align}
{{\mathbf{c}}_k} = \underbrace {{g^c}({g^c}({g^c}({g^c}}_{k{\text{ times}}}( \cdots ),{{\mathbf{x}}_{k - 3}}),{{\mathbf{x}}_{k - 2}}),{{\mathbf{x}}_{k - 1}})
\label{memory_up_ori}
\end{align}
where, at each instant, $g^c$ selectively retains information distilled from historical states and introduces information about new states. In fact, the function nesting in Eq. (\ref{memory_up_ori}) reveals the idea of memory (or hidden variable) iteration in the LSTM \cite{LSTM_ori}. Such a nested form constructs an iterative update of ${\bf c}_k$, which is equivalently expressed as
\begin{align}
{{\mathbf{c}}_k} = g^c\left( {{{\mathbf{c}}_{k - 1}},{{\mathbf{x}}_{k - 1}}} \right)
\label{mem_update}
\end{align}

For the state-dependent observation mismatch error $\Delta _k^h$, we characterize the mapping between it and the state by constructing a function $h^\Delta$:
\begin{align}
\Delta _k^h = h^\Delta \left( {{{\mathbf{x}}_k}} \right)
\label{delta_h}
\end{align}

%The corresponding probabilistic graph model of the transformed model is shown in Fig. \ref{prob_graph}, and s

Through the above function approximation, we discover that the non-Markov system model can be equivalently transformed into a first-order Markov model with memory iteration. In this context, the recursion of memory is affected by the state, while the evolution of the state is also affected by the memory. If ${{\mathbf{c}}_k}$ is considered an extended state, the new state composed of ${{\mathbf{x}}_k}$ and ${{\mathbf{c}}_k}$ remains Markov, demonstrating that non-Markov systems can be equivalently represented using the first-order Markov system with an appropriately extended state dimension. Such a model preserves historical state information while supporting recursive computation. Considering the offline data, the estimation of the model mismatch errors is also transformed into the learning of the functions $f^{\Delta}$, $g^c$ and $h^{\Delta}$ from $\cal D$, which is equivalent to learning the corresponding conditional PDFs. Unlike the traditional Bayesian filtering framework, a triple PDF recursive framework with state, memory, and mismatch is required for this model.

% \begin{figure}[t]
% \centering
% \includegraphics[width=0.8\linewidth]{./fig/Prob_graph.pdf}
% \caption{Probabilistic graph model of joint state-memory iteration}
% \label{prob_graph}
% \end{figure}

\subsection{Bayesian gated framework design}
\begin{figure}[t]
\centering
\includegraphics[width=0.95\linewidth]{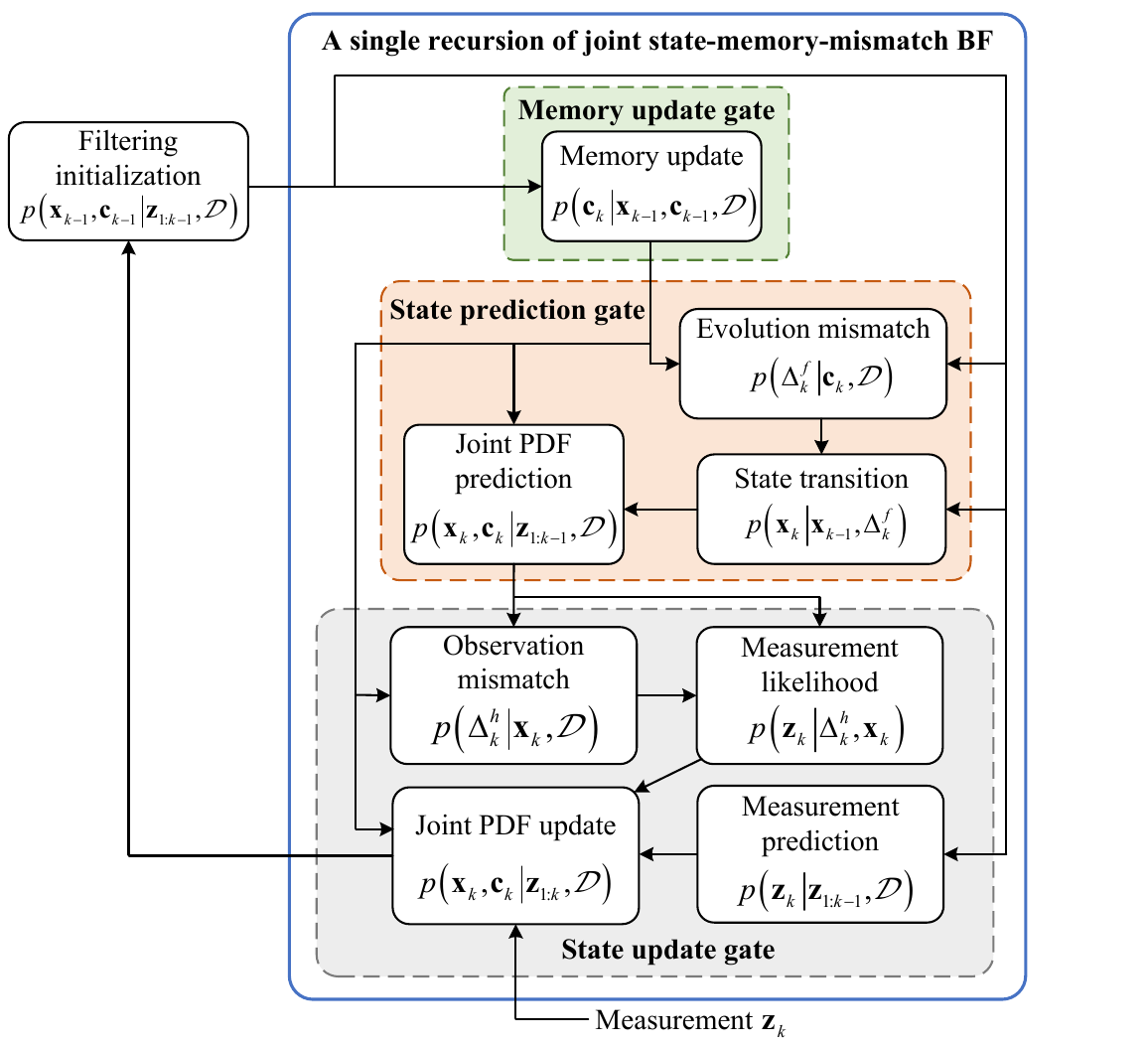}
\caption{Diagram of the joint state-memory-mismatch Bayesian filtering}
\label{prob_framework}
\end{figure}

By introducing the offline data set $\cal D$, we derive the joint state-memory-mismatch Bayesian filtering as shown in Theorem \ref{T}.
\begin{theorem}\label{T}{(Joint state-memory-mismatch Bayesian filtering)}
For the SSM in Eqs. (\ref{sys_error}) - (\ref{delta_f}), (\ref{mem_update}), and (\ref{delta_h}), given the previous measurements ${{\mathbf{z}}_{1:k - 1}}$, the offline data set $\mathcal{D}$, and the previous joint state-memory-mismatch posterior density $p\left( {{{\mathbf{x}}_{k - 1}},{{\mathbf{c}}_{k - 1}}\left| {{{\mathbf{z}}_{1:k - 1}},\mathcal{D}} \right.} \right)$, the joint state-memory density prediction is
\begin{align}
&p\left( {{{\mathbf{x}}_k},{{\mathbf{c}}_k}\left| {{{\mathbf{z}}_{1:k - 1}}}{,\mathcal{D}} \right.} \right) 
\notag \\
&= \iint {P_k^1p\left( {{{\mathbf{x}}_{k - 1}},{{\mathbf{c}}_{k - 1}}\left| {{{\mathbf{z}}_{1:k - 1}},\mathcal{D}} \right.} \right)}d{{\mathbf{x}}_{k - 1}}d{{\mathbf{c}}_{k - 1}}
\label{T1}
\end{align}
with
\begin{align}
P_k^1 =& \int {p\left( {{{\mathbf{x}}_k}\left| {{{\mathbf{x}}_{k - 1}},{\Delta _k^f}} \right.} \right)P_k^2d {\Delta _k^f}}
\label{T2}\\
P_k^2 =& p\left( {\Delta _k^f\left| {{{\mathbf{c}}_k},\mathcal{D}} \right.} \right)p\left( {{{\mathbf{c}}_k}\left| {{{\mathbf{x}}_{k - 1}},{{\mathbf{c}}_{k - 1}},\mathcal{D}} \right.} \right)
\label{T3}
\end{align}

The joint state-memory density update is
\begin{align}
p\left( {{{\mathbf{x}}_k},{{\mathbf{c}}_k}\left| {{{\mathbf{z}}_{1:k}},\mathcal{D}} \right.} \right) =&\frac{{\int {p\left( {{{\mathbf{z}}_k}\left| {\Delta _k^h,{{\mathbf{x}}_k}} \right.} \right)p\left( {\Delta _k^h\left| {{{\mathbf{x}}_k},\mathcal{D}} \right.} \right)d\Delta _k^h} }}{{p\left( {{{\mathbf{z}}_k}\left| {{{\mathbf{z}}_{1:k - 1}},\mathcal{D}} \right.} \right)}} \notag\\
\times& p\left( {{{\mathbf{x}}_k},{{\mathbf{c}}_k}\left| {{{\mathbf{z}}_{1:k - 1}},\mathcal{D}} \right.} \right)
\label{T4}
\end{align}
\label{Bayesian_RNN}
\end{theorem}
\begin{proof}
See Appendix \ref{A}.
\end{proof} 
The probabilistic transfer of the joint state-memory-mismatch BF is shown in Fig. \ref{prob_framework}, and its iteration process contains three gated units: 1) the memory update gate (MUG) determines the current memory based on the previous state, previous memory, and offline data; 2) the state prediction gate (SPG) determines the joint state-memory prediction PDF; 3) the state update gate (SUG) determines the joint update PDF of the state and memory. To get the state prediction and posterior PDFs, the outputs of the state prediction and update gates are computed as
\begin{align}
p\left( {{{\mathbf{x}}_k}\left| {{{\mathbf{z}}_{1:k - 1}},\mathcal{D}} \right.} \right) =& \int {p\left( {{{\mathbf{x}}_k},{{\mathbf{c}}_k}\left| {{{\mathbf{z}}_{1:k - 1}},\mathcal{D}} \right.} \right)d} {{\mathbf{c}}_k}
\label{P_pred_gate}\\
p\left( {{{\mathbf{x}}_k}\left| {{{\mathbf{z}}_{1:k}},\mathcal{D}} \right.} \right) =& \int {p\left( {{{\mathbf{x}}_k},{{\mathbf{c}}_k}\left| {{{\mathbf{z}}_{1:k}},\mathcal{D}} \right.} \right)d} {{\mathbf{c}}_k}
\label{P_update_gate}
\end{align}

Such a gated framework utilizes prior models in the recursive filtering process while integrating offline data for learning the unknown PDFs, including the memory update density $p( {{{\mathbf{c}}_k}\left| {{{\mathbf{x}}_{k - 1}},{{\mathbf{c}}_{k - 1}},\mathcal{D}} \right.} )$, the evolution mismatch density $p( {\Delta _k^f\left| {{{\mathbf{c}}_k},\mathcal{D}} \right.} )$, and the observation mismatch density $p( {\Delta _k^h\left| {{{\mathbf{x}}_k},\mathcal{D}} \right.})$. It is required to design NN modules for the learning of these unknown distributions from offline data, while the analytical implementation of the filtering process can be derived separately in the case that these distributions are known. To fully introduce the proposed gated framework, the filtering implementation is first derived, with the corresponding network module to be designed in the next subsection.

As a general Bayesian filtering framework, the filtering process of the gated framework has various implementation methods, such as Monte Carlo sampling \cite{MC1} and Gaussian approximation \cite{wang2012gaussian}. The gated framework of our EGBRNN is implemented by Gaussian approximation, considering that it is a fast and stable implementation method. In fact, constructing the internal weights or the output of the network in the form of a Gaussian distribution is also widespread in the field of Bayesian NNs \cite{BNN1,BNN2}. The assumptions required to perform the Gaussian approximation are given as follows.

\noindent{Assumption 1.} The process noise ${{\bf{w}}_{k}}$ and the measurement noise ${{\bf{v}}_k}$ obey Gaussian distributions as ${{\bf{w}}_{k}} \sim \mathcal  N\left( {0,{{\bf{Q}}_{k}}} \right)$ and ${{\bf{v}}_{k}} \sim \mathcal  N\left( {0,{{\bf{R}}_{k}}} \right)$, respectively.

\noindent{{Assumption 2.}} The measurement prediction PDF is a Gaussian distribution with first- and second-order moments of ${{\bf{\hat z}}}_{k|k - 1}$ and ${\bf{P}}_{k|k - 1}^{\bf{z}}$, respectively, namely, 
\begin{align}
p\left( {{{\bf{z}}_k}\left| {{{\bf{z}}_{1:k-1}}} \right.} {,\mathcal{D}}\right) = \mathcal  N\left( {{{\bf{z}}_k};{{{\bf{\hat z}}}_{k|k - 1}},{\bf{P}}_{k|k - 1}^{\bf{z}}} \right)
\label{assum_meas}
\end{align}

\noindent {{Assumption 3.}} The state posterior PDF is a Gaussian distribution with first- and second-order moments of ${\bf{\hat x}}_{k|k}$ and ${\bf{P}}_{k|k}$ respectively, namely, 
\begin{align}
{p}\left( {{{\bf{x}}_k}\left| {{{\bf{z}}_{1:k}}} \right.} {,\mathcal{D}}\right) = \mathcal  N\left( {{{\bf{x}}_k};{\bf{\hat x}}_{k|k},{\bf{P}}_{k|k}} \right)
\label{assum_state_up}
\end{align}

\noindent {{Assumption 4.}} ${\bf c}_k$ obeys a Gaussian distribution with first- and second-order moments of $\hat {\bf c}_k$ and ${\bf P}_k^c$, respectively, namely, 
\begin{align}
p\left( {{\bf c}_k\left| {{{\mathbf{x}}_{k - 1}},{{\bf{c}}_{k-1}},\mathcal{D}} \right.} \right) = \mathcal  N\left( {{\bf c}_k;{\bf \hat c}_k,{\bf P}_k^c} \right)
\label{assum_mem}
\end{align}

\noindent {{Assumption 5.}} $\Delta _k^f$ obeys a Gaussian distribution with first- and second-order moments of $\hat \Delta _k^f$ and ${\bf P}_k^f$, respectively; and $\Delta _k^h$ obeys a Gaussian distribution with first- and second-order moments of $\hat \Delta _k^h$ and ${\bf P}_k^h$, respectively, namely,
\begin{align}
p\left( {{\Delta _k^f}\left| {{{\mathbf{c}}_k},\mathcal{D}} \right.} \right) = \mathcal  N\left( {{\Delta _k^f};\hat \Delta _k^f,{\bf P}_k^f} \right)
\label{assum_f}
\\
p\left( {{\Delta _k^h}\left| {{{\mathbf{x}}_k},\mathcal{D}} \right.} \right) = \mathcal  N\left( {{\Delta _k^h};\hat \Delta _k^h,{\bf P}_k^h} \right)
\label{assum_h}
\end{align}

\noindent{\emph{Remark 1:} General Gaussian filtering without mismatch usually assumes that ${{\bf{w}}_{k}}$, ${{\bf{v}}_{k}}$, $p\left( {{{\bf{z}}_k}\left| {{{\bf{z}}_{1:k-1}}} \right.} {,\mathcal{D}}\right)$ and ${p}\left( {{{\bf{x}}_k}\left| {{{\bf{z}}_{1:k}}} \right.} {,\mathcal{D}}\right)$ are Gaussian \cite{wang2012gaussian}, which corresponds to the Assumptions 3, 4, and 5.}

The Gaussian approximation implementation of the filtering process in the proposed gated framework is shown in Theorem 2.
\begin{theorem}\label{TH2}{(Gaussian approximation implementation)} Under Assumptions 1–5, the state and covariance prediction are
\begin{align}
{{{\mathbf{\hat x}}}_{k|k - 1}}=& \int {{{{f}_k}\left( {{{\mathbf{x}}_{k - 1}}} \right)}P_{k-1}^{x+}d{{\mathbf{x}}_{k - 1}}}+ \hat \Delta _k^f
\label{Gau_state_pred}
\\
{{\mathbf{P}}_{k|k - 1}} =& \int {{{f}_k}\left( {{{\mathbf{x}}_{k - 1}}} \right)[f_k\left( {{{\mathbf{x}}_{k - 1}}} \right)]^{\top}P_{k-1}^{x+}d{{\mathbf{x}}_{k - 1}}}  
+ {\mathbf{P}}_k^f \notag \\-& {{{\mathbf{\hat x}}}_{k|k - 1}}{\mathbf{\hat x}}_{k|k - 1}^{\top} + {{\mathbf{Q}}_k}
\label{Gau_cov_pred}
\end{align}
and the state and covariance update are
\begin{align}
{{{\mathbf{\hat x}}}_{k|k}} =& {{{\mathbf{\hat x}}}_{k|k - 1}} + {\mathbf{P}}_{k|k - 1}^{xz}{\left( {{\mathbf{P}}_{k|k - 1}^z} \right)^{ - 1}}\left( {{{\mathbf{z}}_k} - {{{\mathbf{\hat z}}}_{k|k - 1}}} \right)
\label{gau_state_up}\\
{{\mathbf{P}}_{k|k}} =& {{\mathbf{P}}_{k|k - 1}} - {\mathbf{P}}_{k|k - 1}^{xz}{\left( {{\mathbf{P}}_{k|k - 1}^z} \right)^{ - 1}}{\left( {{\mathbf{P}}_{k|k - 1}^{xz}} \right)^{\top}}
\label{gau_cov_up}
\end{align}
where 
\begin{align}
{{{\mathbf{\hat z}}}_{k|k - 1}}=& \int {{{h_k}\left( {{{\mathbf{x}}_k}} \right) }} P_{k}^{x-}d{{\mathbf{x}}_k}+ {\hat \Delta _k^h}
\label{Gau_meas_pred}
\\
{\mathbf{P}}_{k|k - 1}^z=&\int {{h_k}\left( {{{\mathbf{x}}_k}} \right)({{h_k}}\left( {{{\mathbf{x}}_k}} \right))^{\top}P_{k}^{x-}d{{\mathbf{x}}_k}}+  {\bf{P}}_k^{h} 
\notag \\-& {{{\mathbf{\hat z}}}_{k|k - 1}}{\mathbf{\hat z}}_{k|k - 1}^{\top}  + {{\mathbf{R}}_k}
\label{Gau_meas_cov}
\\
{\mathbf{P}}_{k|k - 1}^{xz}=&{\int {{{\mathbf{x}}_k}{({{h_k}}\left( {{{\mathbf{x}}_k}} \right))^{\top}}}}P_{k}^{x-}d{{\mathbf{x}}_k} - {{{\mathbf{\hat x}}}_{k|k - 1}}{\mathbf{\hat z}}_{k|k - 1}^{\top}
\label{Gau_xz_cov}
\\
P_{k-1}^{x+}=&N( {{{\mathbf{x}}_{k - 1}};{{{\mathbf{\hat x}}}_{k - 1|k - 1}},{{\mathbf{P}}_{k - 1|k - 1}}} )\notag
\\
P_{k}^{x-}=&N( {{{\mathbf{x}}_{k}};{{{\mathbf{\hat x}}}_{k|k - 1}},{{\mathbf{P}}_{k|k - 1}}} )
\notag
\end{align}
\end{theorem}
\begin{proof}
See Appendix \ref{B}.
\end{proof} 

In the above implementation process, Gaussian-weighted integrals are crucial. There are various methods for fast calculation of integrals in Gaussian approximate filtering \cite{wang2012gaussian}. Here, we give the computational process under the first-order Taylor expansion. Let ${\bf F}_k$ and ${\bf H}_k$ be the Jacobian matrices of ${f}_k$ and ${h}_k$, respectively. Based on the idea of function approximation used in extended KF (EKF) \cite{EKF_intro}, Eqs. (\ref{Gau_state_pred}) and (\ref{Gau_cov_pred}) are replaced by:
\begin{align}
{{{\mathbf{\hat x}}}_{k|k - 1}} =& {{f}_k}\left( {{{{\mathbf{\hat x}}}_{k - 1|k - 1}}} \right) + \hat \Delta _k^f
\label{ekf_state_pred}\\
{{\mathbf{P}}_{k|k - 1}} =& {{{\mathbf{F}}}_k}{{\mathbf{P}}_{k - 1|k - 1}}{\mathbf{F}}_k^{\top} + {{\mathbf{Q}}_k} + {\mathbf{P}}_k^f
\label{ekf_cov_pred}
\end{align}
and the Eqs. (\ref{Gau_meas_pred})-(\ref{Gau_xz_cov}) are replaced by:
\begin{align}
{{{\mathbf{\hat z}}}_{k|k - 1}} =& {{h}_k}\left( {{{{\mathbf{\hat x}}}_{k|k - 1}}} \right) + \hat \Delta _k^h
\label{ekf_meas_pred}\\
{\mathbf{P}}_{k|k - 1}^z =& {{\mathbf{H}}_k}{{\mathbf{P}}_{k|k - 1}}{\mathbf{H}}_k^{\top}{\text{ + }}{{\mathbf{R}}_k} + {\mathbf{P}}_k^h
\label{ekf_measP_pred}\\
{\mathbf{P}}_{k|k - 1}^{xz} =& {{\mathbf{P}}_{k|k - 1}}{\mathbf{H}}_k^{\top}
\label{ekf_xzP}
\end{align}

Under the Gaussian approximation, the learning of the unknown distributions is transformed into the estimation of their corresponding first- and second-order moments. It is necessary to construct the corresponding mappings by learning from offline data to realize the estimation. In the following, we design the corresponding NN structure for each gated unit in the gated framework and train them as the corresponding estimation mappings via offline data.

\noindent{{\emph{Remark 2: The proposed Bayesian filtering framework's implementation does not strictly require a Gaussian approximation, it merely necessitates a determined PDF. Owing to the substantial computational cost of recursive operations, Bayesian filtering is generally implemented by approximate methods, such as the numerical method or Gaussian approximation. From a practical perspective, the good analytical properties and computational efficiency of the Gaussian approximation render it a suitable method for EGBRNN’s implementation. It streamlines network operations and provides good interpretability. 
Despite the Gaussian assumptions, EGBRNN has some adaptation capability for the non-Gaussian case, which approximates non-Gaussian PDFs by adaptively adjusting key filtering parameters through learning from offline data.}}}

\subsection{Gated network structure design}
\begin{figure*}[t]
\centering
\includegraphics[width=1\linewidth]{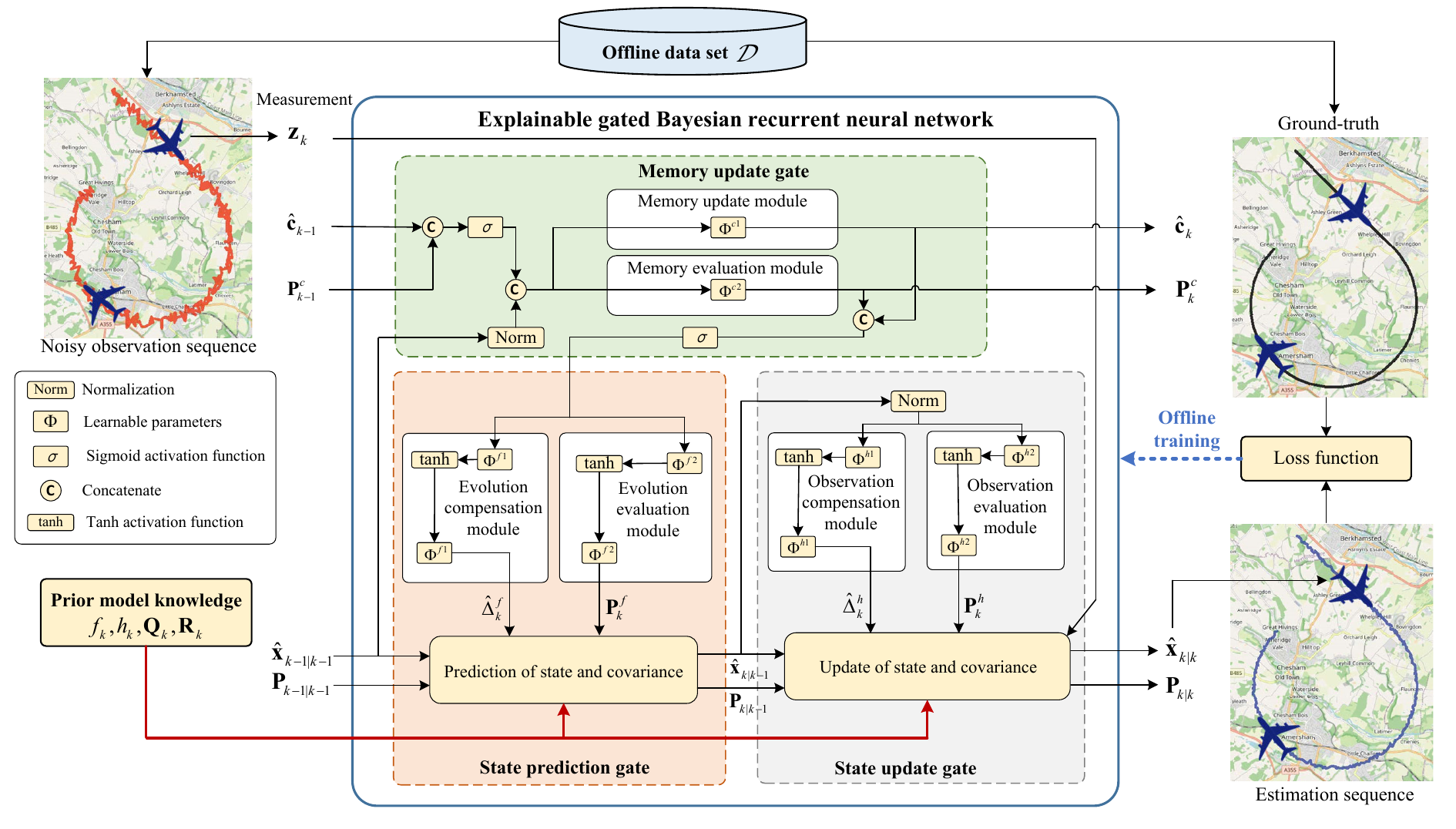}
\caption{\textbf{The overall structure of the EGBRNN and its training process.} The gated structure with the memory update gate, the state prediction gate, and the state update gate is rigorously derived from Bayesian filtering theory and integrates prior model knowledge and offline data efficiently.}
\label{framework}
\end{figure*}
The overall structure of the proposed network is shown in Fig. \ref{framework}, which contains six NN modules in its three gated units for estimating the first- and second-order moments of memory and model mismatch errors, respectively. The prior model information is introduced through the filtering formulas, and the offline data is used for the learning of the NN, while such a structure is rigorously derived based on the BF theory. Benefiting from this, the network modules in each gated unit do not need complex structures. Therefore, we construct the same network with a conventional form for these network modules, i.e., a two-layer NN with the hyperbolic tangent activation function, which is described as a function $F$:
\begin{align}
F\left( {{\bf i},j} \right) = {\omega ^{j2}}\left( {\tanh \left( {{\omega ^{j1}}{\bf i} + {{\mathbf{b}}^{j1}}} \right)} \right) + {{\mathbf{b}}^{j2}}
\label{NN_structure}
\end{align}
where ${\bf i}$ is the input vector and $j$ is the index; $\omega$ and $\bf b$ are the weight and bias of the NN, respectively, and we denote them by a collection of learnable parameters, i.e., ${\omega^{j1}},{{\mathbf{b}^{j1}}},{\omega^{j2}},{{\mathbf{b}^{j2}}} \in {\mathbf{\Phi }^{j}}$; and $\text{tanh}$ denotes the hyperbolic tangent activation function. 

The inputs to the NN modules of each gated unit are described as follows, with descriptions of some of the involved notations provided here: $\sigma$ denotes the sigmoid function,  $con[{\bf a}, {\bf b}]$ denotes the concatenation of vectors $\bf a$ and $\bf b$, $\text{flat}$ denotes the flattening of the matrix into vectors, and $\Psi$ denotes the maximal normalization operation. The normalization, flattening, and other operations we conducted are standard procedures in DL.

\subsubsection{Input to the NN in MUG} In the MUG, NN modules are utilized for the estimation of ${\bf \hat c}_k$ and ${\bf P}_k^c$. Therefore, based on the conditions of PDF $p\left( {{\bf c}_k\left| {{{\mathbf{x}}_{k - 1}},{{\mathbf{c}}_{k-1}},\mathcal{D}} \right.} \right) $, the corresponding input vector is constructed as ${\bf i}_k^c = {\text{con}}\left[ {\sigma \left( {{\text{con}}\left[ {{{{\mathbf{\hat c}}}_{k - 1}},{\text{flat}}\left( {{\mathbf{P}}_{k - 1}^c} \right)} \right]} \right),\Psi\left( {{{{\mathbf{\hat x}}}_{k - 1|k - 1}}} \right)} \right]$.

\subsubsection{Input to the NN in SPG} 
In the SPG, NN modules are utilized for the estimation of $\hat \Delta _k^f$ and ${\mathbf{P}}_{k}^f$. Therefore, based on the conditions of the PDF $p( {{\Delta _k^f}\left| {{{\mathbf{c}}_k},\mathcal{D}} \right.})$, the corresponding input vector is constructed as ${\mathbf{i}}_k^f = {\sigma \left( {{\text{con}}\left[ {{{{\mathbf{\hat c}}}_k},{\text{flat}}\left( {{\mathbf{P}}_k^c} \right)} \right]} \right)}$.

\subsubsection{Input to the NN in SUG}
In the SUG, NN modules are utilized for the estimation of $\hat \Delta _k^h$ and ${\mathbf{P}}_{k}^h$. Therefore, based on the conditions of the PDF $p\left( {{\Delta _k^h}\left| {{{\mathbf{x}}_k},\mathcal{D}} \right.} \right)$, the corresponding input vector is constructed as ${\mathbf{i}}_k^h =  \Psi\left( {{{{\mathbf{\hat x}}}_{k|k - 1}}} \right)$.
\begin{table}[htb!]
\renewcommand{\arraystretch}{1.5}
\centering
\caption{Computation of neural network modules within each gated unit.}
\setlength{\tabcolsep}{5.mm}
\label{NNmapping}
\begin{tabular}{ccccccc}
\hline
                {\textbf{Gate}}  & {\textbf{Module name}}   & {\textbf{Mapping}} \\ \hline
\multirow{2}{*}{MUG} &Memory update  &${{{\mathbf{\hat c}}}_k} = F( {{\mathbf{i}}_k^c,c1} )$  \\
                  &Memory measure& ${\mathbf{P}}_k^c = F( {{\mathbf{i}}_k^c,c2} )$ \\ \hline 
\multirow{2}{*}{SPG} &Evolution compensation  &$\hat \Delta _k^f = F( {{\mathbf{i}}_k^f,f1} )$ \\
                  &Evolution measure  &${\mathbf{P}}_{k}^f=F( {{\mathbf{i}}_k^f,f2} )$  \\ \hline 
\multirow{2}{*}{SUG} & Observation compensation & $\hat \Delta _k^h =F( {{\mathbf{i}}_k^h,h1} )$ \\
                  &Observation measure& ${\mathbf{P}}_{k}^h=F( {{\mathbf{i}}_k^h,h2} )$ \\ \hline 
\end{tabular}
\end{table}

Based on the structures and inputs of the gated units described above, their internal NN modules are constructed as in Tab. \ref{NNmapping}. Obviously, each gated unit has its own specified function, and the errors caused by model mismatches are targeted to be compensated.

Compared to the existing algorithmic frameworks that incorporate gated RNNs like LSTM and GRU into Bayesian filters, EGBRNN is an underlying gated RNN tailored for state estimation and designed based on BF theory. The gated units in EGBRNN possess finely partitioned functionality, rendering them physically interpretable and potentially reducing the demand for learnable parameters, thus diminishing the need for a large amount of training data. Despite the extensive research on task-specific underlying gated RNNs in recent years \cite{FAST_GRNN,AAAI_GRNN,Att_GRNN_TIP,Att_GRNN_SPL}, gated RNNs tailored specifically for state estimation remain largely unexplored, with EGBRNN introducing a novel perspective in this research direction.

\subsection{Algorithm training}

As depicted in Fig. \ref{framework}, in a supervised manner, the gated network within EGBRNN undergoes end-to-end training by extracting the underlying regularities governing state evolution and observation from ${\cal D}$. Despite the fact that the NN computes intermediate parameters instead of directly generating estimated states, we employ the loss function computed from the estimated states and their corresponding ground-truth in ${\cal D}$ to simultaneously train all the learnable parameters within EGBRNN. The specific training process is described as follows.

In line with the commonly used minimum mean square error (MMSE) criterion in BF, we employ the mean square error (MSE) loss function with the addition of L2 regularization for training EGBRNN. Consider the offline data set described in Eq. (\ref{offline_data}), the loss function is described as
\begin{flalign}
{l_{i} }\left ( \Phi \right ) =\frac{1}{K} \sum_{k=1}^{K} \left \| {{\bf \hat x}_{k|k}^{i}\left( \Phi \right) -{\bf x}_{k}^{i}} \right \|^{2}+\tau \left \| \Phi \right \| ^{2}   
\label{loss_i}
\end{flalign}
where $\tau$ is a coefficient used to adjust the proportion of the regular term; $i \in \left\{ {1,2, \cdots ,I} \right\}$ represents the index of the training sample; and $\Phi$ is the learnable parameter set of all six network modules, i.e., $\Phi  = \left\{ {{\Phi ^{c1}},{\Phi ^{c2}},{\Phi ^{f1}},{\Phi ^{f2}},{\Phi ^{h1}},{\Phi ^{h2}}} \right\}$. In Eq. (\ref{loss_i}), the MSE loss directly quantifies the fit of the model to the data, focusing on minimizing the discrepancy between estimation and ground-truth, while the L2 regularization term is added to prevent overfitting by penalizing large weights, thus enhancing the model's generalization capability.

A variation of mini-batch stochastic gradient descent is utilized to optimize $\Phi$. Specifically, for each batch indexed by $n$, we select $J < I$ trajectories denoted as $i_{1}^{n}, i_{2}^{n}, \ldots, i_{J}^{n}$. The mini-batch loss is then computed as follows:
\begin{flalign}
L_n\left (\Phi \right ) =\frac{1}{J} \sum_{j=1}^{J} {l_{i_{j}^{n}}\left ( \Phi \right )  } 
\label{loss_bt}
\end{flalign}

The end-to-end training of the six network modules in EGBRNN is realized by conducting the stochastic gradient descent for gradients of the loss to learnable parameters. Since all operations in the filtering process of the EGBRNN implemented by Gaussian approximation, as in Eqs. (\ref{ekf_state_pred})-(\ref{ekf_xzP}) and (\ref{gau_state_up})-(\ref{gau_cov_up}), are differentiable, these gradients can be computed. Considering complex nonlinear operations may be involved, it is intractable to compute these gradients manually, so we compute them automatically with the auto-differentiation tool \cite{tensorflow}. Given that EGBRNN features a recursive architecture, the backpropagation through time (BPTT) algorithm \cite{bptt} is utilized for its training, which is a technique commonly used in training RNNs. The BPTT algorithm efficiently captures temporal dependencies by unfolding the network over time and computing gradient estimates in both forward and backward directions, making it effective for handling sequential data. Concretely, we utilize shared network parameters to unfold EGBRNN over time and calculate the forward and backward gradient estimations across the network.

\noindent{\emph{Remark 3: 
The spatial placement of the sub-network modules, as well as their inputs and outputs, in the EGBRNN's internal gated structure are derived from Bayesian filtering theory, giving these sub-network modules a well-defined physical meaning. In the forward filtering process, these physically well-defined network modules work together to achieve accurate state estimation. Through end-to-end training, each network module optimizes its learnable parameters based on its placement and role in the overall network. Therefore, these network modules are able to learn the optimal parameters required to perform their specific functions based on their effect on the final state estimation results. The subsequent ablation experiment demonstrates the effectiveness of these network modules' functions.}} 

\section{Experiments}
In this section, the EGBRNN is comprehensively evaluated in various state estimation tasks under unknown or mismatched SSMs. The experiments contain not only real data for two representative applications of state estimation (target tracking and navigation), but also simulated data with non-Markov and nonlinear properties. Through the comparison with various benchmark and SOTA estimation methods, we demonstrate the estimation performance of EGBRNN in non-Markov, non-Gaussian, and nonlinear cases, as well as its efficiency in data dependency, computation time, and parameter utilization. In addition, the effectiveness of model compensation and memory iteration in EGBRNN is also verified. The contents of each experiment are outlined as follows:
\begin{itemize}
\item {\bf {Non-Markov time series estimation.}} This experiment aims to evaluate the non-Markov data processing performance and the dependence on training data volume of EGBRNN. To this end, EGBRNN is compared with several SOTA non-Markov time series processing models, like Transformer \cite{Transformer}, Mamba \cite{MAMBA}, and the alternating direction method of multipliers (ADMM) based non-Markov estimation method \cite{TSP_NOM}, in a non-Markov time series estimation task.

\item {\bf{Landing aircraft tracking.}} This experiment aims to evaluate the model compensation ability of EGBRNN for real-world data with non-Markov features. Here, EGBRNN is compared with representative target tracking methods, like the benchmark tracking algorithm IMM and the SOTA DL-based tracking algorithm DeepMTT \cite{DeepMTT}, in an aircraft landing trajectory tracking task.

\item {\bf{Landing aircraft tracking under glint noise.}} This experiment aims to further evaluate the robustness and scalability of EGBRNN in the non-Gaussian case by tracking landing aircraft under the non-Gaussian glint observation noise.

\item {\bf Odometer-based robot navigation.} This experiment aims to evaluate the estimation performance, computational efficiency, and parameter utilization efficiency of EGBRNN in a real-world navigation task. Under the NCLT dataset \cite{NCLT}, EGBRNN is compared with SOTA state estimation methods such as KalmanNet \cite{KalmanNet} and variational Bayesian based adaptive KF (VBKF) \cite{VBI_huang}.

\item {\bf Lorentz attractor.} This experiment aims to evaluate the estimation performance of EGBRNN in highly nonlinear settings. To this end, EGBRNN tracks the trajectory of the chaotic Lorentz attractor under model mismatch and is compared with several benchmark nonlinear filters such as EKF, UKF, and PF.

\item {\bf Ablation experiment.} Finally, an ablation study is conducted to verify the effectiveness of the model compensation and memory iteration of EGBRNN by separately evaluating the effectiveness of each gated unit in it.
\end{itemize}

Throughout the experiment, we employ two units to quantify estimation errors: MSE in [dB], commonly used in signal processing, is utilized for simulation scenarios to narrow down the range of values; and root MSE (RMSE), a standard metric in target tracking and navigation, is utilized for the real-world data sets. Here, we describe the calculation for these two units. The mean MSE of the estimated state is calculated as
\begin{align}
M^{\bf x} ={{1 \over {K{N}}}\sum\limits_{i = 1}^{{N}} {\sum\limits_{k = 1}^K {\left( {{{\bf{x}}_{i,k}} - {{{\bf{\hat x}}}_{i,k|k}}} \right)^\top{{\left( {{{\bf{x}}_{i,k}} - {{{\bf{\hat x}}}_{i,k|k}}} \right)}}} } }
\end{align}
where $N$ is the number of test samples and $K$ is the time duration. The mean MSE in [dB] is calculated as $10\times {\log _{10}}\left( {{M^{\bf x}}} \right)$ and the mean RMSE is calculated as $\sqrt {M^{\bf x}}$. In addition, the RMSE at each time for a single test sample under ${N_{{\rm{MC}}}}$ times MC simulations is calculated as
\begin{align}
{\rm{R}}^{\bf{x}}_k = \sqrt {{1 \over {{N_{{\rm{MC}}}}}}\sum\limits_{i = 1}^{{N_{{\rm{MC}}}}} {\left( {{{\bf{x}}_{i,k}} - {{{\bf{\hat x}}}_{i,k|k}}} \right)^\top{{\left( {{{\bf{x}}_{i,k}} - {{{\bf{\hat x}}}_{i,k|k}}} \right)}}} }
\label{RMSE_time}
\end{align}

\subsection{Non-Markov time series estimation}
\label{time_series}
In this experiment, a time series dataset is constructed using an autoregressive (AR) model \cite{AR1} to assess the EGBRNN's capability in estimating states within non-Markov evolutionary processes. The AR model, a statistical tool for predicting future time series points based on past values, posits that each series point is a linear combination of its own past values plus a stochastic error term.

\textbf{Experimental setups.} Considering the prevalence of periodic influences from past states in real-world cases, such as the seasonal effects of climate data, we construct a non-Markov AR model to generate the dataset, where the state evolution is periodically influenced by previous states. The SSM used to generate the time series dataset is defined as
\begin{align}
&{x_k} = a{x_{k - 1}} + \sum\limits_{i = 1}^{k - 1} {{b_i}{x_i}}  + {w_k}\label{AR_SSM_sys}\\
&{z_k} = {x_k} + {v_k}
\label{AR_SSM_meas}
\end{align}
where the forgetting factor $a$, set as $0.5$, balances recent and historical state influences; ${b_i} = \varepsilon  \cos \left( {{{2 \rho \pi i} \over {K - 1}}} \right)$ is the periodically modulated weighting coefficient with a scaling factor $\varepsilon$ and a periodic modulation parameter $\rho$, and $K$ is the duration of the time series. The introducing of $\varepsilon$ and $\rho$ allows different periodic modes to be constructed, thus simulating complex non-Markov time series dynamics. Here we set $\varepsilon=0.2$ and $\rho=4$ for the stability of the time series. The process noise and measurement noise are ${w_k} \sim \mathcal  N\left( {0,{\sigma _w^2}} \right)$ and ${v_k} \sim \mathcal  N\left( {0,{\sigma _v^2}} \right)$, respectively. The labeled data is generated by using Eqs. (\ref{AR_SSM_sys}) and (\ref{AR_SSM_meas}) with $K=100$, and the initial state $x_0$ of each sample is randomly sampled from $10\sim 20$ according to a uniform distribution. Three labeled datasets are generated based on the three noise levels $\left\{\sigma _w,\sigma _v\right\} = \left\{ {1,4} \right\}, \left\{ {2,6} \right\}$, and $ \left\{ {4,10} \right\}$. Each dataset contains 704 samples, of which 640 are used for training and 64 for testing.

For the settings of EGBRNN, its nominal state-evolution function is $f_k\left(x_k \right) = ax_k$, where the absence of $b_i$ leads to a model mismatch related to the non-Markov property of the system's evolution; its nominal state-measurement function is $h_k\left(x_k \right) = x_k$, and the exact statistical distribution about the noise is not provided. 
In such a low-dimensional task, it's crucial to streamline the network's scale to prevent overfitting. Hence, we set both the number of hidden layer nodes and the dimension of the memory vector in EGBRNN to $32$.

The comparison methods are described as follows:
1) Transformer: An autoregressive Transformer block with a decoder-only form is used to perform the real-time estimation, and it synthesizes all the previous information through a self-attention mechanism. Specifically, a single Transformer block is employed with 8 self-attention heads and 64 hidden layer nodes, which input the previous estimated state and the current measurement at each frame.
2) Mamba: The SSM-based Mamba can be used directly for real-time estimation, and it remembers previous input information through a hidden vector. Here, we use a single Mamba block as in \cite{MAMBA} with $128$ hidden layer nodes to directly map the $z_k$ to an estimate of the state $\hat x_k$. 3) LSTM: A single-layer LSTM with the same input-output form and hidden node number as Mamba. 4) ADMM: An online optimaztion based non-Markov time series estimation method, which integrates the measurement data and model by iterative optimization based on the ADMM algorithm, and its setup is referenced in \cite{TSP_NOM}. Besides the above methods, we also construct the EGBRNN without memory by masking the MUG, which is a first-order Markov model based method.

\textbf{Experimental results and analysis}
We first evaluate the methods under different noise levels with complete training data available. The mean RMSE estimated over the test data is shown in Tab \ref{mean_RMSE_1D_test}. In comparison to advanced temporal DL models, EGBRNN has a more reasonable network structure for such an estimation task with random noise, which effectively compensates the uncertain model with the learning of non-Markov state evolution, thus achieving lower estimation errors. Without MUG, EGBRNN is unable to effectively compensate for such non-Markov uncertainty, culminating in a pronounced increase in error. The ADMM-based method is an online optimization method without the ability of offline learning, so it is more suitable for estimation tasks without training data. 
\begin{table}[t]
\renewcommand{\arraystretch}{2.15}
\setlength\tabcolsep{4pt}
\begin{center}
\caption{Mean RMSE of the time series estimation on test set.}
\label{mean_RMSE_1D_test}
\setlength{\tabcolsep}{0.8 mm}
\begin{tabular}{ ccccccc }
\hline
 {\makecell{$\sigma _w,\sigma _v$}} & \textbf{\makecell{LSTM}}& \textbf{\makecell{Transformer}}& \textbf{\makecell{Mamba}}& \textbf{\makecell{ADMM}} &{\makecell{{\textbf{EGBRNN}}\\(no MUG)}}& {\makecell{{\textbf{EGBRNN}}\\(full)}}\\
\hline
{\makecell{$1,4$}}&1.93&1.85&2.10&3.15  &2.23  &\textbf{1.65}   \\
\hline
{\makecell{$2,6$}}&3.19&3.07&3.41& 5.16  &3.72  &\textbf{2.78} \\
\hline
{\makecell{$4,8$}}&5.4  &5.1  &5.3  &7.5  & 5.6 & \textbf{4.6} \\
\hline
\end{tabular}
\end{center}
\end{table}

\begin{figure}[t]
\centering
\includegraphics[width=3.4in]{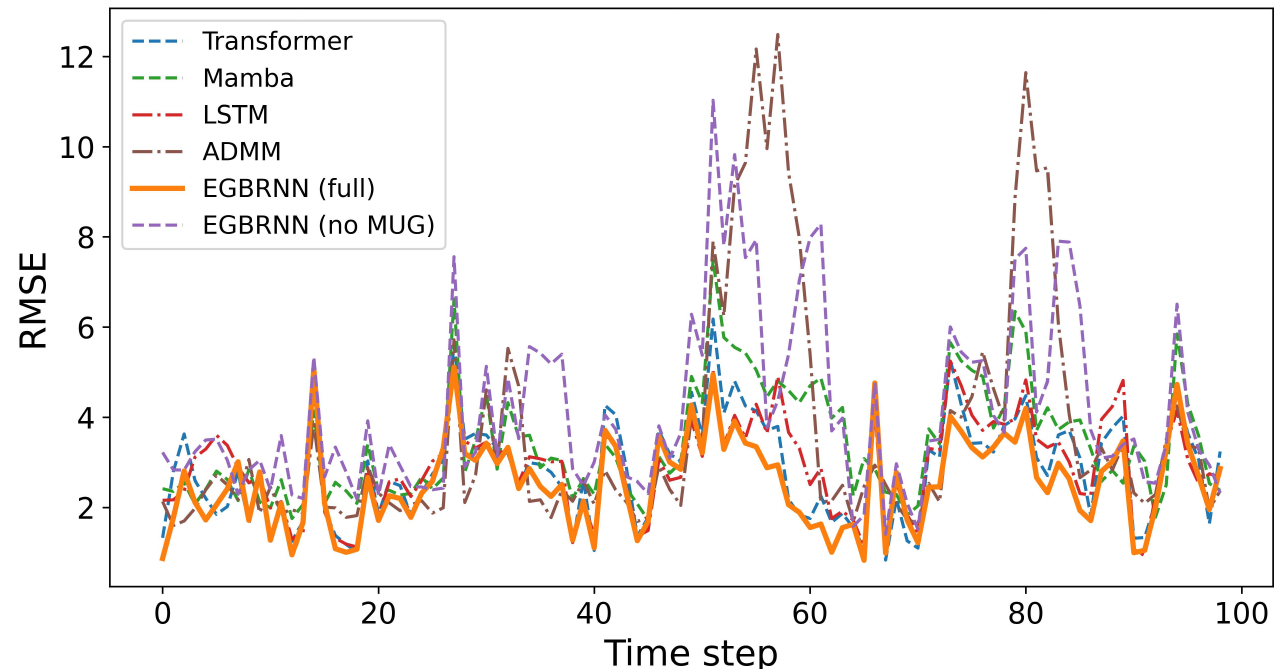}
\caption{RMSE of 100 MC estimation for a single test sample.}
\label{RMSE_1D}
\end{figure}

Fig. \ref{RMSE_1D} further shows the RMSE for a single test sample with $\sigma _w=2$ under 100 times MC estimation, where the measurements for every estimation are randomly generated according to $\sigma _v=6$. It can be seen that EGBRNN has the lowest estimation error in almost all frames. Through the synthesis of previous information to reveal the evolutionary trend, the non-Markov methods generally have low peak errors. The EGBRNN without memory, which uses a first-order Markov model, finds it difficult to capture the temporal information, resulting in a high peak error.

To assess the models' dependency on the volume of training data, they are trained with datasets of smaller sizes, and the noise level of the dataset considered here is $\sigma _w=2$ and $\sigma _v=6$. The mean RMSE on the test set is shown in Table \ref{RMSE_1D_test}. It can be seen that the estimation accuracy of the Transformer and LSTM models significantly degrades with a decrease in the volume of training data. In contrast, Mamba has an SSM-based network architecture that is more effective at learning model information from a small dataset. With only a small volume of training data, EGBRNN still ensures high estimation accuracy, indicating its low dependence on the training data volume. The interpretable internal gated structure ensures its data efficiency in the state estimation task, allowing it to be applied in situations where training data is scarce, and such situations are common in the real world.
\begin{table}[htb]
\renewcommand{\arraystretch}{1.8}
\setlength\tabcolsep{4pt}
\begin{center}
\caption{Mean RMSE of the time series estimation with smaller training data volumes.}
\label{RMSE_1D_test}
\setlength{\tabcolsep}{1.8 mm}
\begin{tabular}{ ccccccc }
\hline
 {\makecell{Number of \\ training samples}} & \textbf{\makecell{LSTM}}& \textbf{\makecell{Transformer}}& \textbf{\makecell{Mamba}}& \textbf{\makecell{EGBRNN}}\\
\hline
{\makecell{$64$}}&4.78&4.57&3.73&\textbf{3.09} \\
{\makecell{$128$}}&4.11&3.92&3.68&\textbf{3.03}\\
{\makecell{$320$}}&4.03&3.23&3.56&\textbf{2.91} \\
\hline
\end{tabular}
\end{center}
\end{table}
\subsection{Landing aircraft tracking}
\label{landing}
In this experiment, our goal is to track the real trajectories of civil aircraft landing at the airport. The tracking of landing aircraft is an important aspect of airspace surveillance, which is critical to flight safety and aircraft scheduling. The landing process of an aircraft is a sequence of typical motions, and the motions in the sequence are influenced by meteorological conditions and air traffic control instructions, which have a certain long-term correlation. The diversity of motion mode combinations and the variable timing of combinations make the correlation difficult to model explicitly, needing to be learned from offline data.

\textbf{Experimental setups.} The trajectory data are from an open dataset in \cite{airdata}, which contains the flight records covering Jan-Mar 2006 in Northern California TRACON. We extracted the trajectories of aircraft landing in the east direction on a designated runway and performed cubic spline interpolation on the trajectories to achieve a fixed sampling interval of $\Delta t = 4$. The trajectory duration is set as $K=200$, and shorter trajectories are discarded or longer ones are truncated accordingly. This resulted in 260 trajectories, with 182 ($70\%$) for training, 52 ($20\%$) for validation, and 26 ($10\%$) for testing. This experiment considers target tracking in a two-dimensional coordinate system, i.e., the state ${\bf{x}}_{k}={\left[ {p_k^x,p_k^y,v_k^x,v_k^y} \right]^{\rm{T}}}$ with position $p$ and velocity $v$. The dataset contains only the ground-truth of the state and does not contain any model and noise information. To construct the complete labeled dataset, the corresponding measurements are generated through a standard radar observation model. Specifically, the measurement is ${{\bf{z}}_{k}}{\rm{ = }}{\left[ {{d_{k}},{\mu _{k}}} \right]^{\top}}$ with the radial distance ${{d_{k}}}$ and the azimuth angle ${\mu _{k}}$, the state-to-measurement function is
\begin{equation}
h({{\bf{x}}_{k}}){\rm{ = }}{\left[ {\sqrt {\left(p_k^x\right)^2 + \left(p_k^y\right)^2} ,\arctan \left( {{{{p_k^y}} \over {{p_k^x}}}} \right)} \right]^{\top}}
\label{radar_h}
\end{equation}and the measurement noise ${{\bf{v}}} = {\left[ {v_d,v_\mu } \right]^\top}$ with ${v_d} \sim \mathcal  N\left( {0,{\sigma _d}^2} \right)$ and ${v_\mu } \sim  \mathcal  N\left( {0,{\sigma _\mu }^2} \right)$. To simulate the radar measurement conditions under different accuracies, four observation noise levels are set to $\left\{ {\sigma_\mu,\sigma_d} \right\}=\left\{ {0.1^\circ,50\text{m}} \right\}$, $\left\{ {0.15^\circ,100\text{m}} \right\}$, $\left\{ {0.2^\circ,100\text{m}} \right\}$, and $\left\{ {0.3^\circ,150\text{m}} \right\}$. The selected radar accuracy levels are based on common operational standards and are designed to test the EGBRNN's adaptability to varying degrees of observation noise, thus providing insights into its robustness and practical applicability.

For the setup of EGBRNN, its observation model is exact since the air traffic control radars are typically cooperative sensors with known accuracy, and its measurement noise covariance is ${\mathbf{R}} = {\rm{diag}}\left( {\sigma _d^2,\sigma _\mu ^2} \right)$. Considering the difficulty of accurately modeling the motion of aircraft, the nominal motion model is set as a constant velocity (CV) model, namely, the nominal state evolution function is
\begin{align}
    {f_k\left({\bf{x}}_{k-1}\right)}&=\begin{bmatrix}
   1 & 0 & {\Delta t} & 0  \cr
   0 & 1 & 0 & {\Delta t}  \cr
   0 & 0 & 1 & 0  \cr
   0 & 0 & 0 & 1  \cr
\end{bmatrix}{\bf{x}}_{k-1}
\label{CV_model}
\end{align}
with the sampling interval $\Delta t = 4$, and the prior process noise covariance is set as ${\bf Q} = {\rm{diag}}\left( {{q^2},{q^2},{q^2},{q^2}} \right)$ with $q^2=10$. This experiment has a higher dimension compared to the one-dimensional time series, so we appropriately increase the learnable parameter to set both the number of hidden layer nodes and the dimension of the memory vector in EGBRNN to $128$.

The comparison methods involved in this experiment are described as follows. 1) DeepMTT \cite{DeepMTT}: A maneuvering target tracking algorithm with a bi-directional LSTM network to compensate for the residual of the unscented Kalman filter (UKF) in sequence-to-sequence form, and exhibits the multi-order Markov property determined by the sequence length. The setting of DeepMTT is referenced in \cite{DeepMTT}, and it is used in the form of a frame-by-frame sliding window with a window length of 5 to satisfy the real-time tracking task. 2) IMM: The base filter of the IMM is UKF, and its model set includes one CV model and two coordinate turn (CT) models with $\omega=\pm 3^\circ/s$, where $-3^\circ/s \sim 3^\circ/s$ is the normal range of turn rates for a civil aircraft. 3) Mamba: the input of Mamba is the transformation of ${\bf z}_k$ to values in the Cartesian coordinate system, and its output is the estimated state. It has 256 hidden layer nodes, and the other settings are the same as in Section \ref{time_series}. 4) EGBRNN without MUG: Its settings are the same as Section \ref{time_series}. For these comparative algorithms, we ensure that the noise settings align with those of the EGBRNN to maintain consistency in evaluation. Except for Mamba, which is unable to utilize the prior noise information.
\begin{figure}[t]
\centering
\includegraphics[width=3.3in]{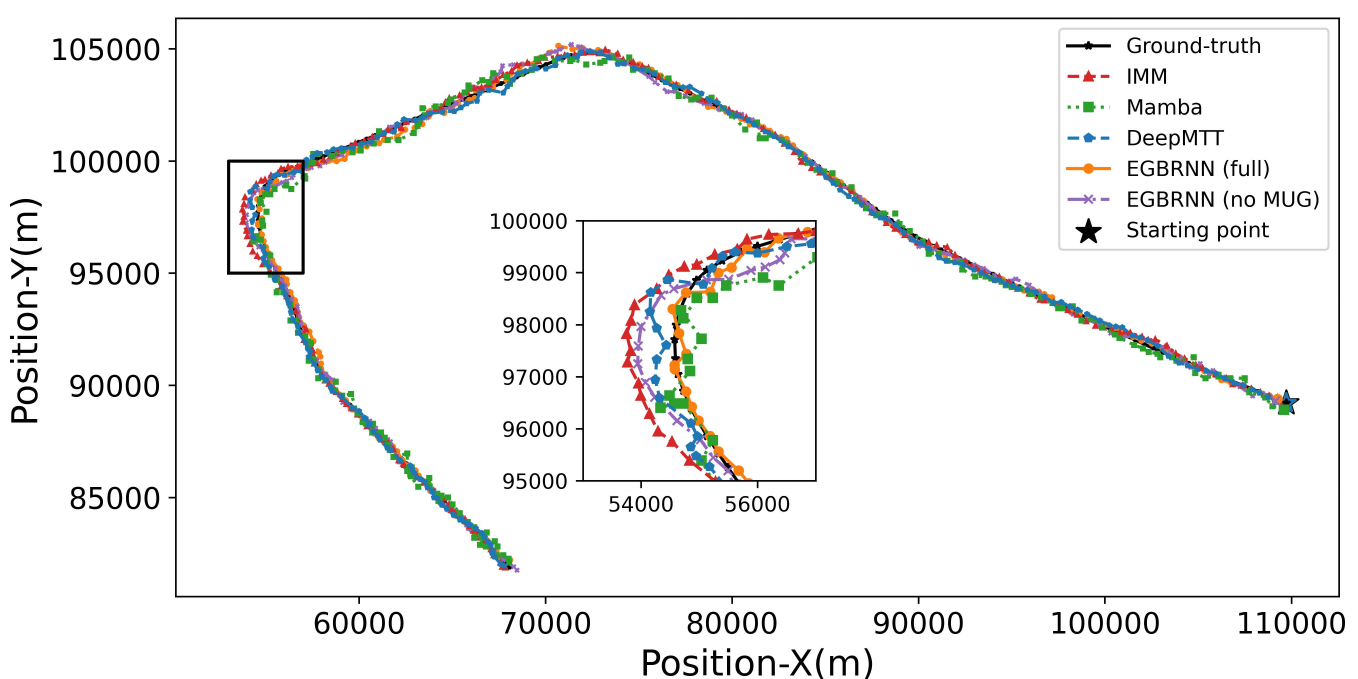}
\caption{Tracked trajectories of each method for a test sample.}
\label{Air_single_track}
\end{figure}

\begin{figure}[t]
\centering
\includegraphics[width=3.3in]{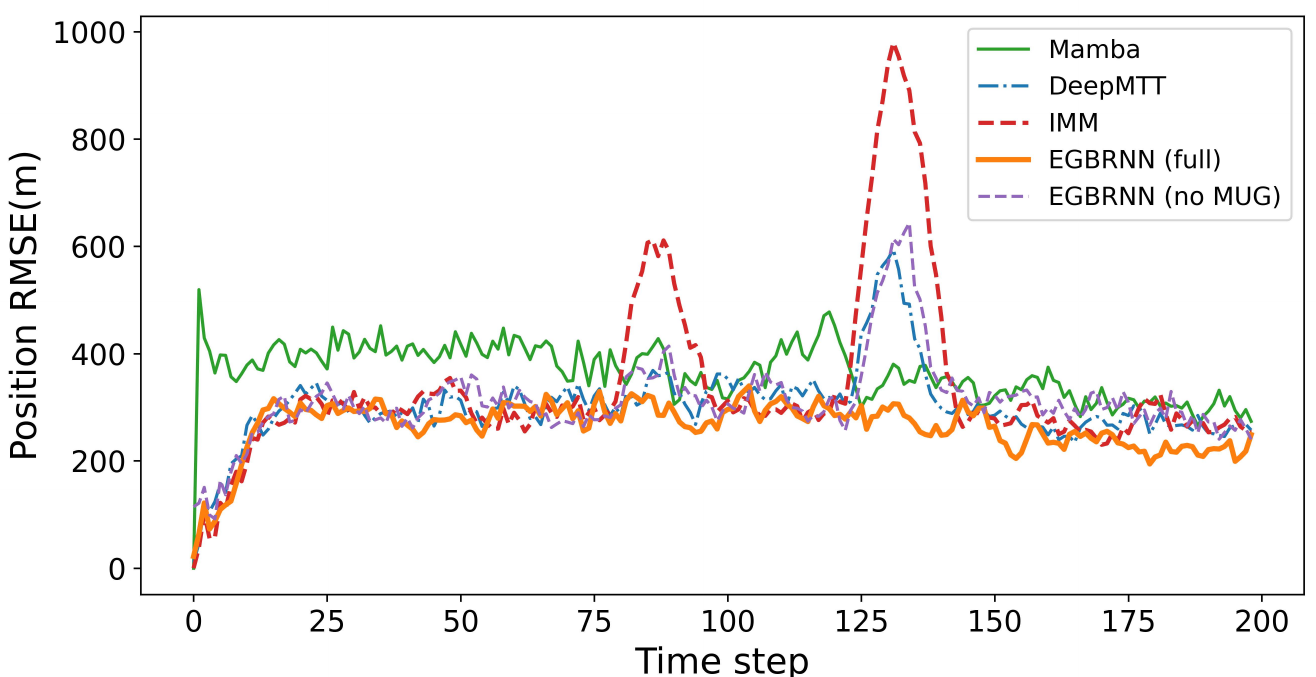}
\caption{Position RMSE (m) of 100 MC tracking for a test sample.}
\label{Air_RMSE_pos}
\end{figure}

\textbf{Experimental results and analysis.} We first performed 100 times MC tracking for each method on a test sample, with the measurement noise being  $\{\sigma_\mu,\sigma_d\}  = \{  0.3^\circ,150\text{m}\} $.  
Fig. \ref{Air_single_track} shows the tracking trajectories for this test sample. It can be seen that the motions of the target contain multiple modes (straight line, turning with varying radii, etc.), and EGBRNN provides accurate tracking during target maneuvering (see zoomed-in section in the figure). Fig. \ref{Air_RMSE_pos} shows the position RMSE of each frame for this sample. EGBRNN exhibits the lowest steady-state error, with virtually no peak error. With the memory mechanism, it learns the non-Markov state evolution law from offline data, which makes it effective at capturing the evolutionary trends of the motion and suppressing peak errors. Mamba also has the ability to capture non-Markov information, but has a high overall error as it fails to introduce the prior model. DeepMTT synthesizes the state in the last five frames, with a peak error higher than the above non-Markov methods, and lower than that of EGBRNN without memory, which uses a first-order Markov model. IMM, which also uses a first-order Markov model, lacks the ability to learn from offline data and exhibits the highest peak error.
\begin{table}[htb!]
\renewcommand{\arraystretch}{2.}
\setlength\tabcolsep{4pt}
\begin{center}
\caption{Mean position RMSE (m) of the landing aircraft tracking on test set.}
\label{aircraft_result_pos}
\setlength{\tabcolsep}{1.5 mm}
\begin{tabular}{ ccccccc }
\hline
{$\sigma_\mu,\sigma_d$} & \textbf{\makecell{IMM}}& \textbf{\makecell{DeepMTT}}& \textbf{\makecell{Mamba}}&{\makecell{{\textbf{EGBRNN}}\\(no MUG)}}& {\makecell{{\textbf{EGBRNN}}\\(full)}}\\
\hline
{\makecell{$ 0.1^\circ,50\text{m} $}}& 151.4&132.9&177.6&139.5&\textbf{124.1} \\
\hline
{\makecell{$ 0.15^\circ,100\text{m} $}}&212.4&182.1&219.6 &191.3&\textbf{171.5}\\
\hline
{\makecell{$ 0.2^\circ,100\text{m} $}}&264.7&234.1&282.4& 241.8&\textbf{205.7} \\
\hline
{\makecell{$ 0.3^\circ,150\text{m} $}}&359.6&310.4&385.7 &318.2&\textbf{266.8}\\
\hline
\end{tabular}
\end{center}
\end{table}
\begin{table}[htb!]
\renewcommand{\arraystretch}{2.}
\setlength\tabcolsep{4pt}
\begin{center}
\caption{Mean velocity RMSE (m/s) of the landing aircraft tracking on test set.}
\label{aircraft_result_vel}
\setlength{\tabcolsep}{1.5 mm}
\begin{tabular}{ ccccccc }
\hline
{$\sigma_\mu,\sigma_d$} & \textbf{\makecell{IMM}}& \textbf{\makecell{DeepMTT}}& \textbf{\makecell{Mamba}}&{\makecell{{\textbf{EGBRNN}}\\(no MUG)}}& {\makecell{{\textbf{EGBRNN}}\\(full)}}\\
\hline
{\makecell{$ 0.1^\circ,50\text{m} $}}&15.6&14.8&37.8&15.3&\textbf{14.2} \\
\hline
{\makecell{$ 0.15^\circ,100\text{m} $}}&17.5&16.2&37.4&16.8 &\textbf{15.1}\\
\hline
{\makecell{$ 0.2^\circ,100\text{m} $}}& 20.2&17.6&38.1&18.9&\textbf{16.5} \\
\hline
{\makecell{$ 0.3^\circ,150\text{m} $}}&22.7&18.8&37.1&19.6&\textbf{16.9}\\
\hline
\end{tabular}
\end{center}
\end{table}

We next statistics the mean RMSE of all test data under various radar accuracy settings of $\sigma _\mu$ and $\sigma _d$, and the corresponding mean position and velocity RMSE are shown in Tabs. \ref{aircraft_result_pos} and \ref{aircraft_result_vel}, respectively. It can be seen that EGBRNN has the lowest tracking error in both position and velocity. In comparison to the MB method, EGBRNN is capable of extracting implicit state evolution modes from offline data. Compared with the pure DL method Mamba, EGBRNN is able to reasonably introduce prior model knowledge relying on its interpretable computational framework, which is essential for complex state evolution in practice. Compared to DeepMTT with a combinatorial structure, the internal structure of EGBRNN is tailored to state estimation under model mismatches, which adequately compensates for tracking errors caused by model mismatches.

\subsection{Landing aircraft tracking under glint noise}
To further evaluate the robustness and scalability of the EGBRNN in the non-Gaussian case, we use it to track the aircraft trajectory data in Section \ref{landing} under the non-Gaussian observation noise. In radar systems, target glint can lead to measurement noise exhibiting non-Gaussian characteristics, often termed glint noise, as described in \cite{glint_1993}. The glint noise is Gaussian-like around the mean and transitions to a non-Gaussian, long-tailed distribution in its tail regions. These tail regions contain outlier data points, which are indicative of spikes caused by glint \cite{glint_2010}.

\textbf{Experimental setups.}  We refer to \cite{glint_2010} to model the glint observation noise as a mixture of a Gaussian and a Laplacian noise. Specifically, for component $j$ of the measurement noise $\bf v$ in the landing trajectory tracking experiment, the glint noise is defined as
\begin{equation}
{v_j} = \left( {1 - {\tau _j}} \right)\mathcal  N\left( {0,\sigma _j^2} \right) + {\tau _j}{\cal L}\left( {0,{\beta _j}} \right)
\label{glint_model}
\end{equation}
where $\tau _j$ is the glint probability, and ${\cal L}\left( {0,{\beta _j}} \right)$ is the Laplacian PDF with zero-mean and the scale parameter $\beta _j$. Considering $\tau _j=0.2$, two glint levels of $\beta _j = 2 \times \sigma _j$ and $\beta _j = 5 \times \sigma _j$ are set to generate the non-Gaussian measurements of the dataset in Section \ref{landing}. Under the radar noise setting of $[\sigma_\mu,\sigma_d] = [ 0.3^\circ,150\text{m}]$, two new non-Gaussian labeled datasets are generated, and the division of the training, validation, and test sets is the same as earlier. Here, EGBRNN is compared with the same comparison methods in Section \ref{landing}, and each method retains the previous settings. Note that ${\beta _j}$ is not provided to the tracking methods, which means that the distribution of the measurement noise is mismatched.

\begin{figure}[t]
\centering
\includegraphics[width=3.3in]{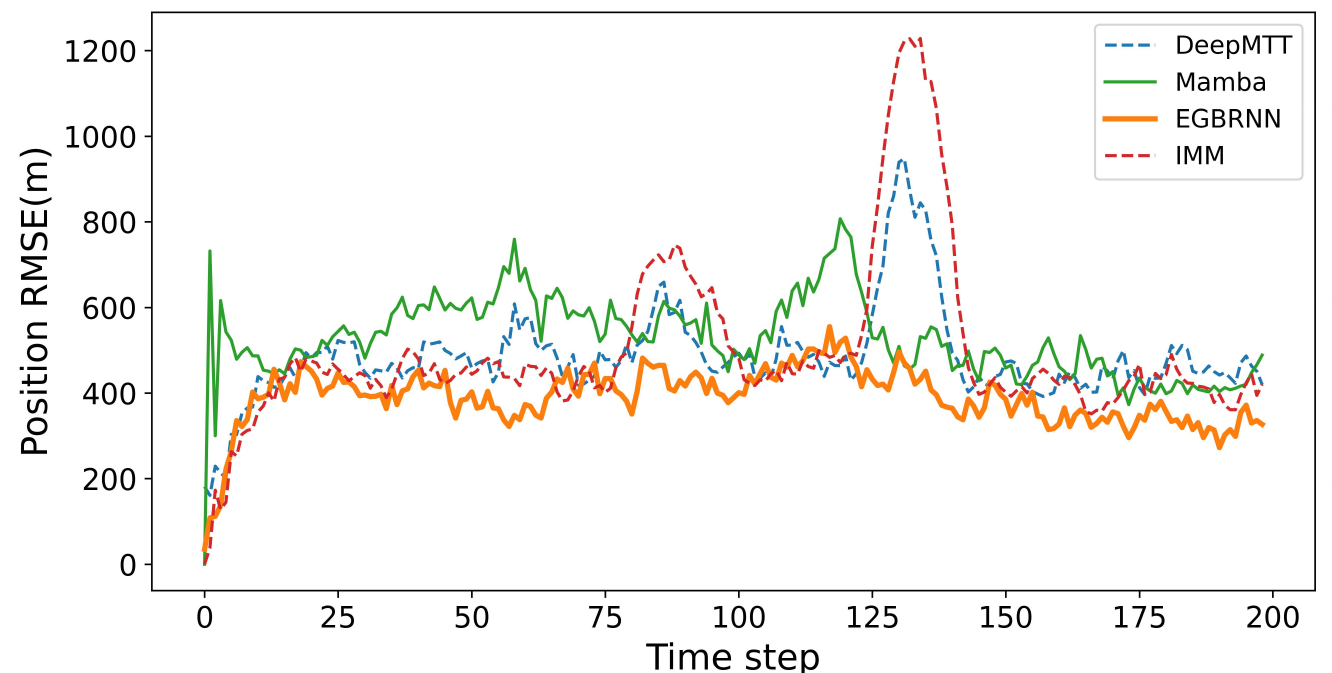}
\caption{Position RMSE of 100 MC tracking under glint noise.}
\label{Air_RMSE_pos_glint}
\end{figure}

\textbf{Experimental results and analysis} Tabs. \ref{aircraft_glint_pos} and \ref{aircraft_glint_vel} show the mean position and velocity RMSE of each method on the test set, respectively. It can be seen that EGBRNN is still effective in improving the tracking accuracy in the non-Gaussian case. By learning from offline data, EGBRNN, implemented by the Gaussian approximation, is able to approximate the non-Gaussian distribution to some extent and achieve stable tracking. The same test sample as in Section \ref{landing} is selected for 100 times MC tracking, and the RMSE is shown in Fig. \ref{Air_RMSE_pos_glint}. The RMSE curve of the EGBRNN has a similar trend to that in the case of Gaussian observation noise, and it effectively suppresses the peak error, which suggests that it has a stable model learning capability in non-Gaussian glint noise.
\begin{table}[htp!]
\renewcommand{\arraystretch}{1.8}
\setlength\tabcolsep{4pt}
\begin{center}
\caption{Mean position RMSE (m) of the landing aircraft tracking under glint noise.}
\label{aircraft_glint_pos}
\setlength{\tabcolsep}{3.4mm}
\begin{tabular}{ cccccc }
\hline
{$\beta_j$} & \textbf{\makecell{IMM}}& \textbf{\makecell{DeepMTT}}& \textbf{\makecell{Mamba}}& \textbf{\makecell{EGBRNN}}\\
\hline
{\makecell{$ 2 \times \sigma_j$}}&389.0&326.2&408.8& \textbf{269.5} \\
\hline
{\makecell{$5 \times \sigma_j$}}&524.3&497.0&556.5 &\textbf{410.1}\\
\hline
\end{tabular}
\end{center}
\end{table}
\begin{table}[htp!]
\renewcommand{\arraystretch}{1.8}
\setlength\tabcolsep{4pt}
\begin{center}
\caption{Mean velocity RMSE (m/s) of the landing aircraft tracking under glint noise.}
\label{aircraft_glint_vel}
\setlength{\tabcolsep}{3.4mm}
\begin{tabular}{ ccccccc }
\hline
{$\beta_j$} & \textbf{\makecell{IMM}}& \textbf{\makecell{DeepMTT}}& \textbf{\makecell{Mamba}}& \textbf{\makecell{EGBRNN}}\\
\hline
{\makecell{$2 \times \sigma_j$}}&23.7&20.4&38.6&\textbf{18.1} \\
\hline
{\makecell{$5 \times \sigma_j$}}&25.7&23.8&39.3&\textbf{20.3}\\
\hline
\end{tabular}
\end{center}
\end{table}
\subsection{Odometer-based robot navigation}
\label{odo_nav}
This experiment evaluates the performance of EGBRNN on a real navigation task. The labeled trajectory data is from the NCLT data set, and our specific goal is to track the two-dimensional position using only odometry data, which means that the observations come from noisy sensor readings while modeling information is not available in the dataset.
Therefore, we need to set the prior nominal model for EGBRNN, but it may be mismatched: it is difficult to accurately characterize the complex motion by a nominal model, and the sensor is affected by perturbation or drift in practical usage.

\textbf{Experimental setups.} The experimental setups refer to \cite{KalmanNet}. In the NCLT data set, the trajectory at 2012-01-22 with outliers removed is segmented into 48 trajectories, each containing a sequence of ground-truth positions and corresponding noisy velocity readings with a duration $K=100$ and a sampling frequency of $1\rm{Hz}$. In these trajectories, we set 34 for training ($71\%$), 9 for validation ($19\%$), and 5 for testing ($10\%$). The state vector is ${\bf{x}}_{k}={\left[ {p_k^x,p_k^y,v_k^x,v_k^y} \right]^{\rm{T}}}$ with ${\left[ {p_k^x,p_k^y} \right]^\top}$ and ${\left[ {v_k^x,v_k^y} \right]^{\top}}$ being the position and velocity, respectively, and the measurement is the noisy velocity. 

For the setup of EGBRNN, the model information is not provided in the dataset, we need to manually set up the nominal model used for navigation. The nominal state-evolution function is a CV model as in Eq. (\ref{CV_model}) with the sampling interval $\Delta t = 1$, and the process noise covariance is set as ${\bf Q} = {\rm{diag}}\left( {{0.01},{0.01},{0.01},{0.01}} \right)$. A simple linear model is used to model the observations, i.e., the nominal state-measurement function is
\begin{align}
h_k\left( {\bf{x}}_k \right) = \left[ {\begin{array}{*{20}{c}}
  0&{\begin{array}{*{20}{c}}
  {\begin{array}{*{20}{c}}
  0&1 
\end{array}}&0 
\end{array}} \\ 
  0&{\begin{array}{*{20}{c}}
  {\begin{array}{*{20}{c}}
  0&0 
\end{array}}&1 
\end{array}} 
\end{array}} \right] {\bf{x}}_k
\label{meas_linear}
\end{align}
and the measurement noise covariance is ${\mathbf{R}} = {\rm{diag}}\left( {1,1} \right)$. The number of hidden layer nodes in the EGBRNN is set to $64$ and the dimension of its memory vector ${\bf {c}}_k$ is also $64$.

In this experiment, we compare EGBRNN with four types of representative methods, namely, the MB KF, the pure DL method LSTM and Transformer, the SOTA online Bayesian optimization method VBKF, and the SOTA combination method KalmanNet. In these methods, KF uses a CV motion model, and its noise setting is determined by a grid search; LSTM has a single-layer structure with $128$ hidden layer nodes and directly maps velocity measurements to positions; the Transformer has the same structure as that in Section \ref{time_series}; and the settings of VBKF and KalmanNet are references to \cite{VBI_huang} and \cite{KalmanNet}, respectively.

\begin{figure}[t]
\centering
\includegraphics[width=3.in]{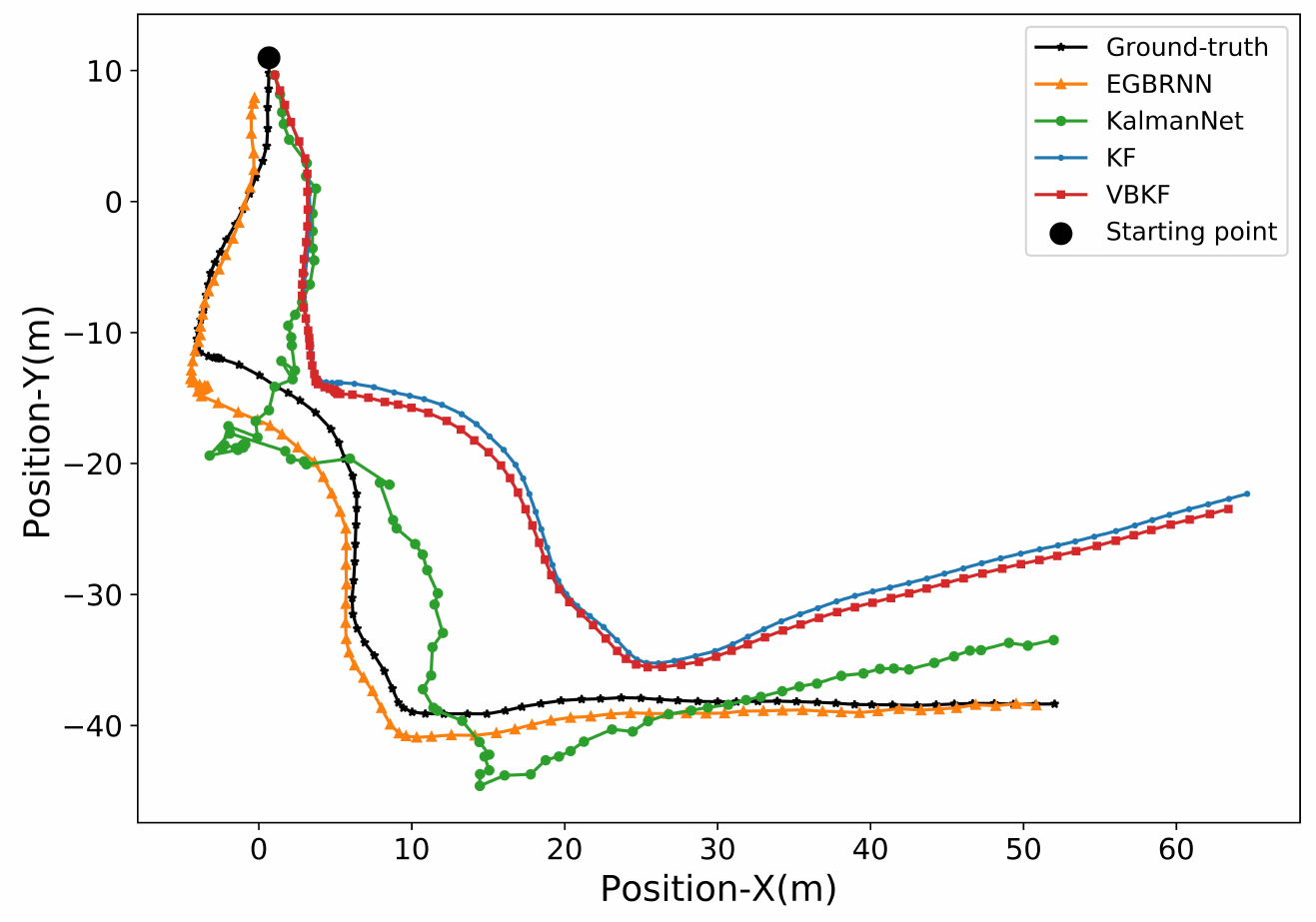}
\caption{Localization trajectories on a test sample in NCLT dataset.}
\label{NCLT_res3}
\end{figure}

\textbf{Experimental results and analysis.} Tab. \ref{NCLT_result} shows the localization RMSE of each method on the test set. It can be seen that EGBRNN achieves the lowest localization error in such a challenging real-world navigation task. This is supported by its explicitly functional gated unit to rationally compensate for model mismatches. KF without learning capability has the highest localization error, and VBKF slightly outperforms KF by online inference to the covariance. KalmanNet performs offline learning while introducing prior model knowledge, but it only modifies the filter gain, making it incapable of compensating for model mismatches effectively. The localization trajectories of a single test sample are shown in Fig. \ref{NCLT_res3}, where we can see that EGBRNN has a high localization accuracy with low termination error. 
\begin{table}[htb]
\renewcommand{\arraystretch}{1.7}
\begin{center}
\caption{Localization Results for the NCLT test data.}
\label{NCLT_result}
\setlength{\tabcolsep}{0.5mm}
\begin{tabular}{ ccccccc }
\hline
\textbf{\makecell{Method}}& \textbf{\makecell{KF}}& \textbf{\makecell{VBKF}}&\textbf{\makecell{LSTM}}&\textbf{\makecell{Transformer}}&\textbf{\makecell{KalmanNet}}& \textbf{\makecell{EGBRNN}}\\
\hline
\textbf{\makecell{RMSE\\(m)}} & 14.48&13.37&98.81&44.65&12.09&\textbf{2.25} \\
\hline
\textbf{\makecell{ Run time\\(sec)}}& 0.006 &0.134 &0.009 &0.101 &0.115 &0.012  \\
\hline
\end{tabular}
\end{center}
\end{table}

The accurate localization of the EGBRNN does not come at the cost of slow computational inference, which is demonstrated by the average inference time (without parallel computation) shown in Tab. \ref{NCLT_result}. The inference times for each method are measured on the same platform -- the Ubuntu 20.04 operating system with a Linux kernel; CPU: AMD EPYC 7T83 64-Core Processor; GPU: NVIDIA GeForce RTX 4090. It can be seen that the EGBRNN achieves fast inference, which is only slightly slower than the basic models with high localization error, i.e., LSTM and KF.

\begin{figure}[t]
\centering
\includegraphics[width=3.in]{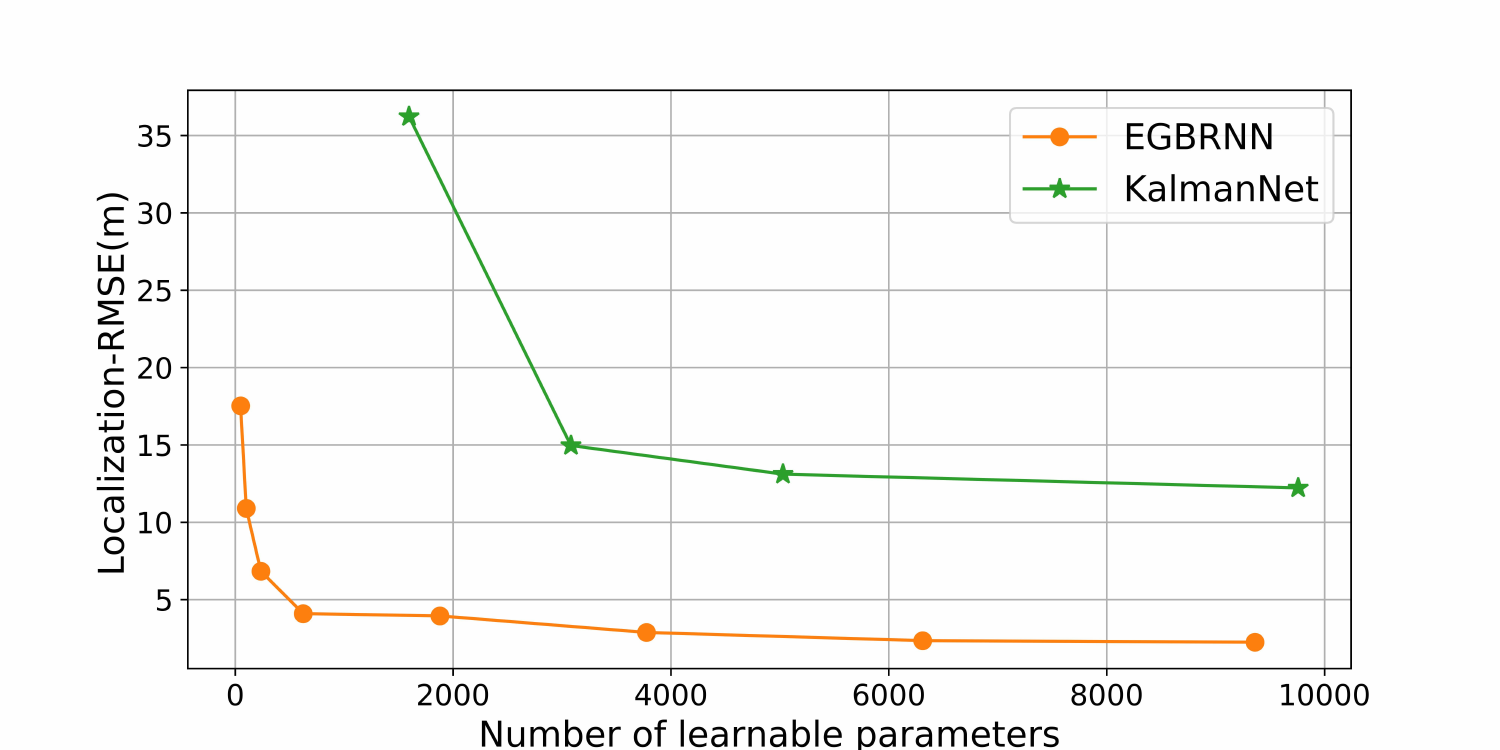}
\caption{The mean localization RMSE (y-axis) of EGBRNN and KalmanNet on the test set under different numbers of learnable parameters (x-axis).}
\label{para_num}
\end{figure}

Fig. \ref{para_num} shows the mean localization RMSE results for EGBRNN and KalmanNet with varying numbers of learnable parameters to further demonstrate its efficiency for parameter utilization. It can be seen that EGBRNN achieves lower localization errors despite having fewer learnable parameters. The EGBRNN is an underlying gated RNN with a computationally efficient gated structure, which allows it to excel in state estimation tasks while maintaining simplicity. In fact, in such a task with very limited training samples, the small parameter scale of EGBRNN confers a significant advantage.

\subsection{Lorenz Attractor}
The Lorenz attractor is a chaotic system generated by a set of differential equations used to simulate atmospheric convection, characterized by ever-changing three-dimensional trajectories. In this scenario, the tracking ability of the EGBRNN in complex and highly nonlinear settings under model mismatch is validated.

\textbf{Experimental setups.} The Lorentz attractor scenario here has the same parameter settings as in \cite{KalmanNet} and \cite{TGRS_tracking}. The Lorenz attractor is mathematically described by a system of three coupled ordinary differential equations, and the noiseless state-evolution of the continuous time process ${{\mathbf{x}}_t} = {\left[ {x_t^1,x_t^2,x_t^3} \right]^\top}$ is 
\begin{flalign}
\frac{{\partial {{\mathbf{x}}_t}}}{{\partial t}} = {\mathbf{A}}\left( {{{\mathbf{x}}_t}} \right){{\mathbf{x}}_t},{\mathbf{A}}\left( {{{\mathbf{x}}_t}} \right) = \left[ {\begin{array}{*{20}{c}}
  { - \sigma }&\sigma &0 \\ 
  \beta &{ - 1}&{ - {x_t^1}} \\ 
  0&{{x_t^1}}&{ - \rho } 
\end{array}} \right]
\end{flalign}
where $\sigma=10$, $\rho=8/3$, and $\beta=28$ are parameters governing the system's dynamics. We follow the processing in \cite{KalmanNet} to get a discrete-time state-evolution model, and the resulting discrete-time evolution process is ${{\mathbf{x}}_{k + 1}} = {\mathbf{F}}\left( {{{\mathbf{x}}_k}} \right){{\mathbf{x}}_k}$, where ${{\mathbf{x}}_k} = {\left[ {x_k^1,x_k^2,x_k^3} \right]^{\top}}$ is a discrete-time state vector, and the discrete-time state-evolution matrix ${\mathbf{F}}\left( {{{\mathbf{x}}_k}} \right)$ is obtained by the Taylor series expansion and a finite series approximation (with $J$ coefficients). The total frame of the generated trajectory is $2000$, of which we train only the initial 100 frames and test the whole trajectory. Three trajectories are generated by considering three measurement noise levels of ${1}/{{{r^2}}}\left[ \text{dB} \right] \in \left\{ { - 10,0,10} \right\}$, and none of these trajectories add process noise. 

The estimation is performed under a model mismatch. The real observation model is an identity matrix, while the nominal state-measurement function of EGBRNN is a rotation with $\theta = 10^\circ$ for it, in the same way as in \cite{KalmanNet}. The nominal state-evolution model of EGBRNN is $f_k\left({{\bf{x}}_k}\right)={\mathbf{F}}\left( {{{\mathbf{x}}_k}} \right)$ with $J=1$ while the actual data is generated by ${\mathbf{F}}\left( {{{\mathbf{x}}_k}} \right)$ with $J = 5$. The number of hidden layer nodes in EGBRNN is set to $64$ and the dimension of its memory vector ${\bf {c}}_k$ is also $64$. Under different noise levels, EGBRNN is compared with several benchmark nonlinear filters, i.e., the EKF, UKF, and particle filter (PF), and their noise settings are determined by grid search.
\begin{figure}[t]
\centering
\includegraphics[width=3in]{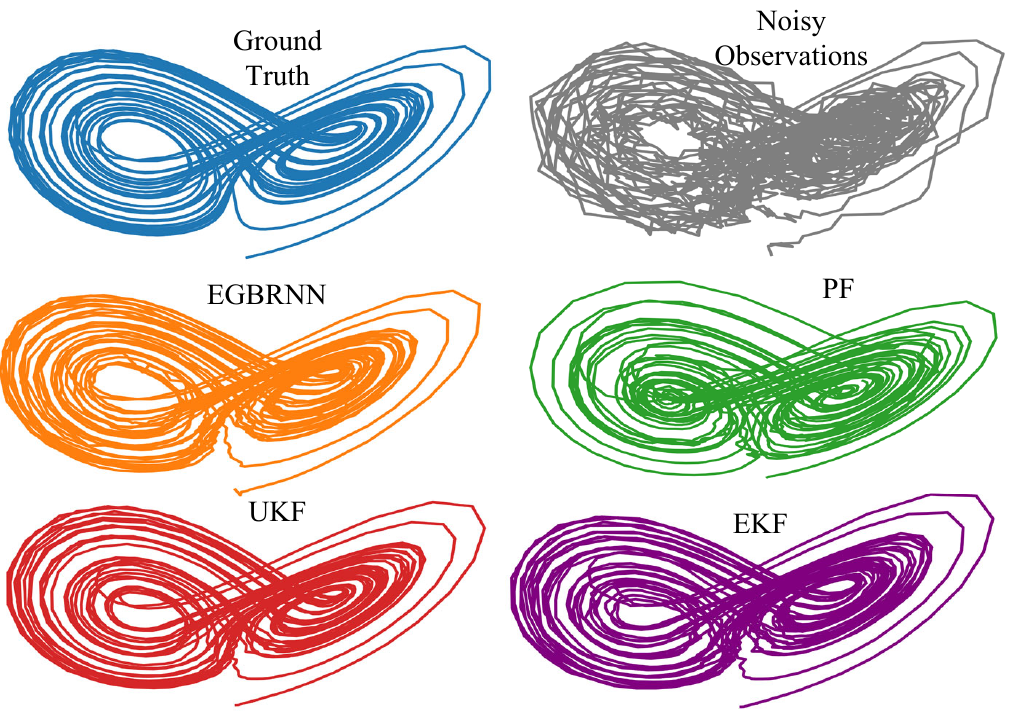}
\caption{Tracking trajectories of Lorenz attractor under model mismatch}
\label{Lorenz_track}
\end{figure}

\textbf{Experimental results and analysis.} The mean MSE in [dB] of each method is shown in Tab. \ref{Lorenz_noF_result}. EGBRNN exhibits the lowest estimation error, highlighting its capacity to significantly improve performance in scenarios with severe nonlinearities through offline learning. Other MB nonlinear filters all degrade in performance under model mismatch, especially under lower measurement noise. The tracking trajectories of each method for a noise level of $1/r^2 = 0 [\text{dB}]$ are shown in Fig.  \ref{Lorenz_track}, with the trajectory estimated by EGBRNN being more closely fitting the ground-truth.
\begin{table}[htp]
\renewcommand{\arraystretch}{1.4}
\begin{center}
\caption{Mean MSE[dB] of tracking the Lorenz attractor under model mismatches}
\label{Lorenz_noF_result}
\setlength{\tabcolsep}{4.mm}
\begin{tabular}{ ccccc }
\hline
{$1/{r^2} [\text{dB}]$} & \textbf{\makecell{EKF}}& \textbf{\makecell{UKF}}& \textbf{\makecell{PF}}&
\textbf{\makecell{EGBRNN}}\\
\hline
{\makecell{-10}}& 5.65&5.25& 1.12&\textbf{0.69} \\
{\makecell{0}}&2.32&2.04& -2.02&\textbf{-7.78} \\
{\makecell{10}}&4.84&4.42 &-1.67 &\textbf{-13.94} \\ 
\hline
\end{tabular}
\end{center}
\end{table}

\subsection{Ablation Experiment}
\label{ablation}
To evaluate the effectiveness of the memory mechanism and model compensation, an ablation study is conducted for EGBRNN in the odometer-based navigation experiment mentioned in Section \ref{odo_nav}. 

\textbf{Experimental setups.} With the same data setups in Section \ref{odo_nav}, we set four schemes by masking the different gated units in EGBRNN to evaluate the effectiveness of each gated unit: 1) full EGBRNN with three gated units; 2) EGBRNN without the MUG; 3) EGBRNN without the SPG; and 4) EGBRNN without the SUG. In Scheme 2, the input of the SPG does not contain the memory but only the state estimation of the previous instant.
\begin{figure}[t]
\centering
\includegraphics[width=3.3in]{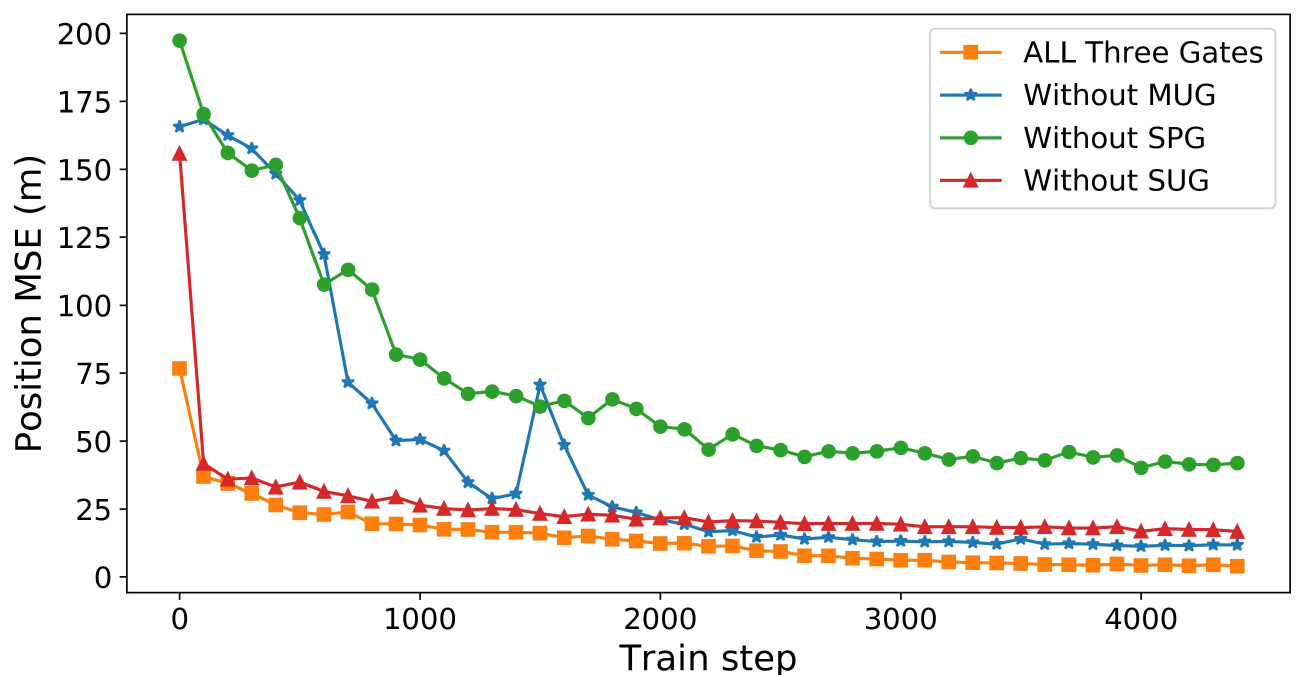}
\caption{\textbf{Training loss on the validation set.} The absence of each gate affects the convergence speed and final convergence result of EGBRNN training.}
\label{Loss_val}
\end{figure}
\begin{figure}[t]
\centering
\includegraphics[width=3.in]{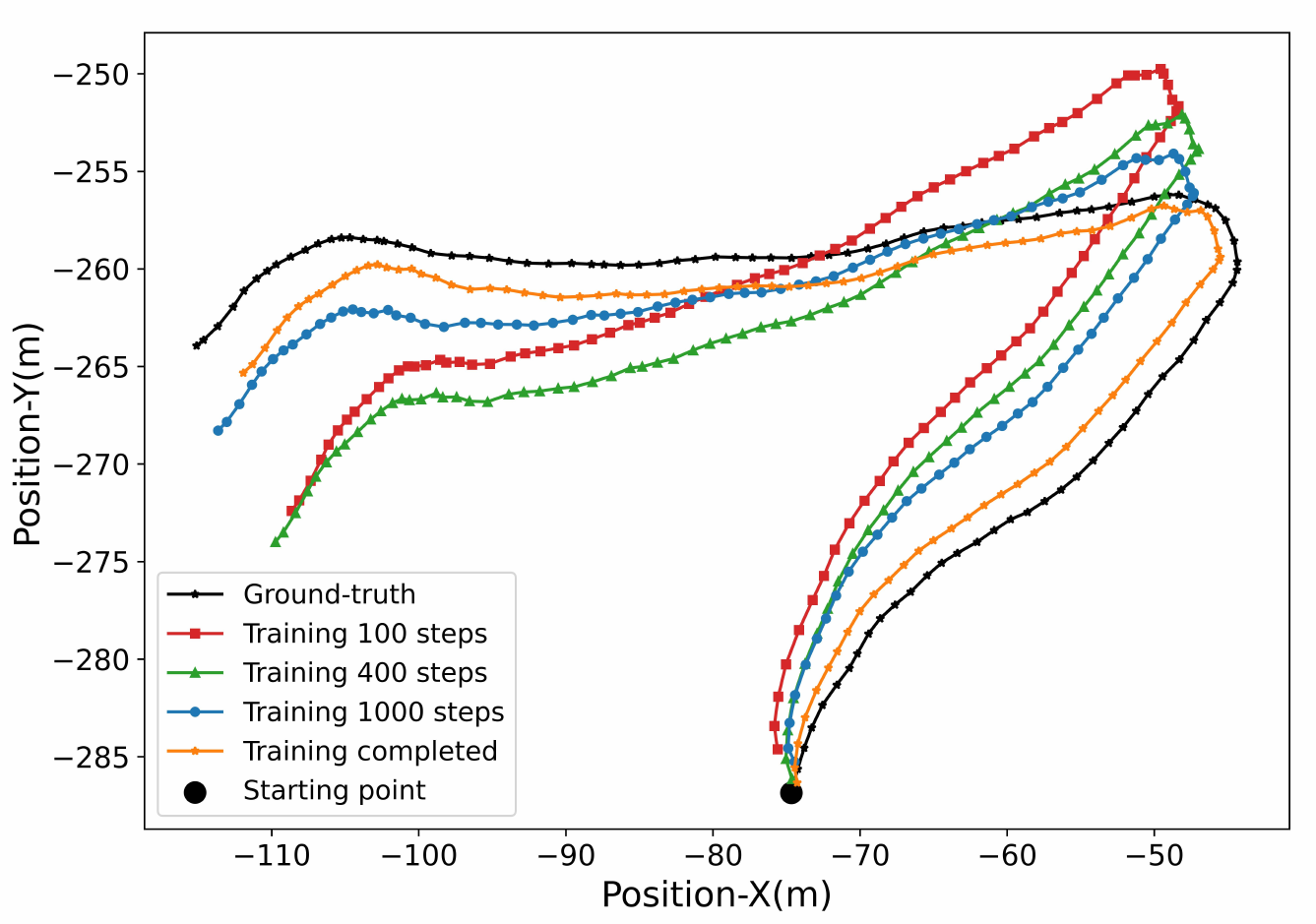}
\caption{Localization trajectories on validation data in different training stages.}
\label{NCLT_train_process}
\end{figure}

\textbf{Experimental results and analysis}
Fig. \ref{Loss_val} depicts the convergence of localization errors on the validation set during the training process of the four schemes. EGBRNN with a full-gated structure exhibits the quickest convergence to achieve the lowest localization error. Although the scheme without the MUG converges quickly, it is constrained by the first-order Markov property, resulting in greater localization error. The scheme without the SUG cannot compensate for the observation mismatch, leading to slow convergence and a high loss. The scheme without the SPG trains the worst since learning the dynamic model is the most critical of state estimation. We further show the localization trajectories of the EGBRNN for a single sample on the validation set at different training stages, as shown in Fig. \ref{NCLT_train_process}. As the training progresses, the EGBRNN continuously improves the localization accuracy, which indicates the correctness of the learning process.
\begin{table}[htb]
\renewcommand{\arraystretch}{1.5}
\begin{center}
\caption{Mean localization RMSE (m) of different ablation schemes on test data.}
\label{NCLY_abl_t}
\setlength{\tabcolsep}{2.mm}
\begin{tabular}{ ccccccc }
\hline
\textbf{\makecell{Scheme}}& \textbf{\makecell{No MUG}}& \textbf{\makecell{No SPG}}&\textbf{\makecell{No SUG}}&\textbf{\makecell{Full EGBRNN}}\\
\hline
\textbf{\makecell{RMSE}} & 3.77&9.54&4.91 &\textbf{2.25} \\
\hline
\end{tabular}
\end{center}
\end{table}

The mean localization RMSE of the four schemes on the test set is shown in Tab. \ref{NCLY_abl_t}. It can be seen that masking out each gated unit will increase localization errors, reflecting the effectiveness of each gated unit in EGBRNN. 
The gated structure of EGBRNN is derived from Bayesian filtering theory, which gives each of its gated units a clearly defined physical meaning. The explicit physical meaning brings them different functions, including memory iteration, dynamic model compensation, and observation compensation. And masking out any of the gated units will result in the absence of a particular function, thus decreasing the estimation performance.

\section{CONCLUSION}
In this paper, an explainable gated Bayesian RNN is specifically designed for non-Markov state estimation under model mismatch, i.e., EGBRNN. The gated structure of EGBRNN is designed by introducing the memory mechanism and offline data into the BF theory, which provides good interpretability, low complexity, and data efficiency. Each gated unit in the gated framework has its own explicit functionality, which effectively compensates for the errors resulting from model mismatches by integrating prior models and offline data. For computational efficiency, the Gaussian approximation is used to implement the filtering process of the gated framework, along with the development of an internal neural network structure and the corresponding end-to-end training method. The EGBRNN has been verified by both simulations and real-world data sets. It is demonstrated that the proposed algorithm outperforms the state-of-the-art DL-based filtering methods and could achieve high estimation accuracy with a small number of learnable parameters.

\appendices
\section{Proof of the joint state-memory-mismatch BF}
\label{A}
Denote that $p\left( {{{\bf{x}}_{1:K}},{{\bf{c}}_{1:K}}\left| {{{\bf{z}}_{1:K}}},{\mathcal{D}} \right.} \right)$ is the joint posterior density of 1 to $K$ time states and memories. Expanding this density under the Bayesian theorem as
\begin{align}
&p\left( {{{\bf{x}}_{1:K}},{{\bf{c}}_{1:K}}\left| {{{\bf{z}}_{1:K}}},{\mathcal{D}} \right.} \right)\notag
\\&= p\left( {{{\bf{x}}_{1:K}},{{\bf{c}}_{1:K}}\left| {{{\bf{z}}_K},{{\bf{z}}_{1:K - 1}}},{\mathcal{D}} \right.} \right) \notag
\\&= \frac{{p\left( {{{\mathbf{z}}_K}\left| {{{\mathbf{x}}_{1:K}},{{\mathbf{c}}_{1:K}},{{\mathbf{z}}_{1:K - 1}}} \right.} \right)p\left( {{{\mathbf{x}}_{1:K}},{{\mathbf{c}}_{1:K}}\left| {{{\mathbf{z}}_{1:K - 1}},\mathcal{D}} \right.} \right)}}{{p\left( {{{\mathbf{z}}_K}\left| {{{\mathbf{z}}_{1:K - 1}},\mathcal{D}} \right.} \right)}}
\label{appendA_1}
\end{align}

According to Eq. (\ref{delta_h}), the measurement is independent of the previous state given the current state. In addition, the state information contained in previous measurements has been accurately characterized by the current state. Then we have
\begin{align}
p\left( {{{\bf{z}}_K}\left| {{{\bf{x}}_{1:K}},{{\bf{c}}_{1:K}},{{\bf{z}}_{1:K - 1}}},{\mathcal{D}} \right.} \right) = p\left( {{{\bf{z}}_K}\left| {{{\bf{x}}_K}},{\mathcal{D}} \right.} \right)
\label{appendA_2}
\end{align}

According to the total probability formula, we have
\begin{align}
&p\left( {{{\bf{x}}_{1:K}},{{\bf{c}}_{1:K}}\left| {{{\bf{z}}_{1:K - 1}}},{\mathcal{D}} \right.} \right)  \notag \\
&= p\left( {{{\bf{x}}_K},{{\bf{c}}_{K}}\left| {{{\bf{x}}_{K - 1}},{{\bf{c}}_{K - 1}},{{\bf{z}}_{1:K - 1}}},{\mathcal{D}} \right.} \right)  \notag
\\ &\times p\left( {{{\bf{x}}_{1:K - 1}},{{\bf{c}}_{1:K - 1}}\left| {{{\bf{z}}_{1:K - 1}}},{\mathcal{D}} \right.} \right)
\label{appendA_3}
\end{align}

Thus, (\ref{appendA_1}) is further written as
\begin{align}
&p\left( {{{\bf{x}}_{1:K}},{{\bf{c}}_{1:K}}\left| {{{\bf{z}}_{1:K}}} \right.}{,\mathcal{D}} \right)  \notag
\\ & =  \frac{{p\left( {{{\bf{z}}_K}\left| {{{\bf{x}}_K}},{\mathcal{D}} \right.} \right)p\left( {{{\bf{x}}_K},{{\bf{c}}_{K}}\left| {{{\bf{x}}_{K - 1}},{{\bf{c}}_{K - 1}},{{\bf{z}}_{1:K - 1}}},{\mathcal{D}} \right.} \right)}}{{p\left( {{{\mathbf{z}}_K}\left| {{{\mathbf{z}}_{1:K - 1}},\mathcal{D}} \right.} \right)}}  \notag
\\ &\times p\left( {{{\bf{x}}_{1:K - 1}},{{\bf{c}}_{1:K - 1}}\left| {{{\bf{z}}_{1:K - 1}}},{\mathcal{D}} \right.} \right)
\label{appendA_4}
\end{align}

For Eq. (\ref{appendA_4}), the conditional independence expressed by Eqs. (\ref{sys_error}) and (\ref{mem_update}) leads to
\begin{align}
&p\left( {{{\bf{x}}_K},{{\bf{c}}_{K}}\left| {{{\bf{x}}_{K - 1}},{{\bf{c}}_{K -1}},{{\bf{z}}_{1:K - 1}}},{\mathcal{D}} \right.} \right) \notag
\\ &= p\left( {{{\bf{x}}_K}\left| {{{\bf{c}}_{K }},{{\bf{x}}_{K - 1}},{{\bf{z}}_{1:K - 1}}},{\mathcal{D}} \right.} \right)  
\notag
\\ &\times p\left( {{{\bf{c}}_{K }}\left| {{{\bf{x}}_{K - 1}},{{\bf{c}}_{K - 1}},{{\bf{z}}_{1:K - 1}}},{\mathcal{D}} \right.} \right) \notag\\&= \int {p\left( {{{\bf{x}}_K}\left| {{{\bf{x}}_{K - 1}},{ \Delta _k^f}} \right.},{\mathcal{D}} \right)} P_{K - 1}^3d{ \Delta _k^f}
\label{appendA_5}
\end{align}
For $P_{K - 1}^3$, considering the transfer relations in Eq. (\ref{delta_f}) and Eq. (\ref{sys_error}) yields
\begin{align}
P_{K - 1}^3 =& p\left( {{\Delta _k^f},{{\bf{c}}_{K}}\left| {{{\bf{x}}_{K - 1}},{{\bf{c}}_{K - 1}},{{\bf{z}}_{1:K - 1}}},{\mathcal{D}} \right.} \right) \notag\\=& p\left( {{ \Delta _k^f}\left| {{{\bf{c}}_{K }}},{\mathcal{D}} \right.} \right)  p\left( {{{\bf{c}}_{K}}\left| {{{\bf{c}}_{K - 1}},{{\bf{x}}_{K - 1}},{{\bf{z}}_{1:K - 1}}},{\mathcal{D}} \right.} \right)
\label{appendA_6}
\end{align}

Based on Chapman-Kolmogorov equations for Eq. (\ref{appendA_5}), the joint state-memory prediction density $p\left( {{{\bf{x}}_k},{{\bf{c}}_k}\left| {{{\bf{z}}_{1:k - 1}}},{\mathcal{D}} \right.} \right)$ can be obtained as Eqs. (\ref{T1}) - (\ref{T3}). 

Based on the Bayesian theorem, the joint state-memory update density can be obtained as
\begin{align}
p\left( {{{\bf{x}}_k},{{\bf{c}}_k}\left| {{{\bf{z}}_{1:k}}},{\mathcal{D}} \right.} \right)  =\frac{{p\left( {{{\bf{z}}_k}\left| {{{\bf{x}}_k}},{\mathcal{D}} \right.} \right)p\left( {{{\bf{x}}_k},{{\bf{c}}_k}\left| {{{\bf{z}}_{1:k - 1}}} ,{\mathcal{D}}\right.} \right)}}{{p\left( {{{\mathbf{z}}_k}\left| {{{\mathbf{z}}_{1:k - 1}},\mathcal{D}} \right.} \right)}} 
\end{align}

From Eq.(\ref{meas_error}) it follows that ${\bf z}_k$ is only related to the current state ${\bf{x}}_k$ and not to the memory ${{\bf{c}}_k}$, thus
\begin{align}
p\left( {{{\bf{z}}_k}\left| {{{\bf{x}}_k}},{\mathcal{D}} \right.} \right)=&
p\left( {{{\bf{z}}_k}\left| {{{\bf{x}}_k}},{\mathcal{D}} \right.} \right)
\notag \\
=&{\int {p\left( {{{\mathbf{z}}_k}\left| {\Delta _k^h,{{\mathbf{x}}_k}} \right.} \right)p\left( {\Delta _k^h\left| {{{\mathbf{x}}_k},\mathcal{D}} \right.} \right)d\Delta _k^h} }
\label{up_proven}
\end{align}

\section{Proof of the Gaussian approximation implementation}
\label{B}
The mean of the state prediction is calculated as:
\begin{align}
{{{\mathbf{\hat x}}}_{k|k - 1}} =& {\text{E}}\left[ {{{\mathbf{x}}_k}\left| {{{\mathbf{z}}_{1:k - 1}},\mathcal{D}} \right.} \right] 
\notag \\
=& E\left[ {{{\bar f}_k}\left( {{{\mathbf{x}}_{k - 1}}} \right) + \Delta _k^f + {{\mathbf{w}}_k}\left| {{{\mathbf{z}}_{1:k - 1}},\mathcal{D}} \right.} \right]
\notag \\
=& \iiiint {{{ {\left[ {{{\bar f}_k}\left( {{{\mathbf{x}}_{k - 1}}} \right) + \Delta _k^f} \right]P_k^1d\Delta _k^fd{{\mathbf{c}}_k}d{{\mathbf{c}}_{k - 1}}d{{\mathbf{x}}_{k - 1}}} } } }
\label{Gau1}
\end{align}
with
\begin{align}
P_k^1 =& p\left( {\Delta _k^f\left| {{{\mathbf{c}}_k},\mathcal{D}} \right.} \right)p\left( {{{\mathbf{c}}_k}\left| {{{\mathbf{x}}_{k - 1}},{{\mathbf{c}}_{k - 1}},\mathcal{D}} \right.} \right)
\notag \\
\times& p\left( {{{\mathbf{x}}_{k - 1}},{{\mathbf{c}}_{k - 1}}\left| {{{\mathbf{z}}_{1:k - 1}},\mathcal{D}} \right.} \right)
\label{Gau2}
\end{align}
and
\begin{align}
&\int {p\left( {{{\mathbf{x}}_{k - 1}},{{\mathbf{c}}_{k - 1}}\left| {{{\mathbf{z}}_{1:k - 1}},\mathcal{D}} \right.} \right)d} {{\mathbf{c}}_{k - 1}} 
\notag\\
&= p\left( {{{\mathbf{x}}_{k - 1}}\left| {{{\mathbf{z}}_{1:k - 1}},\mathcal{D}} \right.} \right) 
\notag \\
&= \mathcal  N\left( {{{\mathbf{x}}_{k - 1}};{{{\mathbf{\hat x}}}_{k - 1|k - 1}},{{\mathbf{P}}_{k - 1|k - 1}}} \right)
\label{Gau3}
\end{align}

Substituting $p( {{{\mathbf{c}}_k}\left| {{{\mathbf{x}}_{k - 1}},{{\mathbf{c}}_{k - 1}},\mathcal{D}} \right.} )=\mathcal  N\left( {{{\mathbf{c}}_k};{{{\mathbf{\hat c}}}_k},{\mathbf{P}}_k^c} \right)$ and $p( {{\Delta _k^f}\left| {{{\mathbf{c}}_k},\mathcal{D}} \right.})=N( {{\Delta _k^f};\hat \Delta _k^f,{\bf P}_k^f} )$ into Eq. (\ref{Gau2}), Eq. (\ref{Gau1}) can be calculated as Eq. (\ref{Gau_state_pred}).

Considering the prediction residual is ${\varepsilon^x  _k} = {{\mathbf{x}}_k} - {{{\mathbf{\hat x}}}_{k|k - 1}}$, the state prediction covariance is calculated as
\begin{align}
{{\mathbf{P}}_{k|k - 1}} =& {\text{E}}\left[ {{\varepsilon^x _k}{{{\varepsilon^x _k}}^{\top}}\left| {{{\mathbf{z}}_{1:k - 1}},\mathcal{D}} \right.} \right] 
\notag \\
=& \iiiint {M_k^fP_k^1d\Delta _k^fd{{\mathbf{c}}_k}d{{\mathbf{c}}_{k - 1}}d{{\mathbf{x}}_{k - 1}}}
\label{Gau7}
\end{align}
with
\begin{align}
M_k^f = \left( {{{f}_k}\left( {{{\mathbf{x}}_{k - 1}}} \right) + \Delta _k^f + {{\mathbf{w}}_k} - {{{\mathbf{\hat x}}}_{k|k - 1}}} \right){\left(  \cdot  \right)^{\top}}
\label{Gau8}
\end{align}

Substituting $N( {{{\mathbf{c}}_k};{{{\mathbf{\hat c}}}_k},{\mathbf{P}}_k^c} )$ and $N( {{\Delta _k^f};\hat \Delta _k^f,{\bf P}_k^f} )$ into Eq. (\ref{Gau7}) while considering Eq. (\ref{Gau3}), then we have the covariance prediction as in Eq. (\ref{Gau_cov_pred}).

In the state update stage, we compute the state posterior according to the Bayesian 
rule:
\begin{align}
p\left( {{{\mathbf{x}}_k}\left| {{{\mathbf{z}}_{1:k}},\mathcal{D}} \right.} \right) = \frac{{p\left( {{{\mathbf{x}}_k},{{\mathbf{z}}_k}\left| {{{\mathbf{z}}_{1:k - 1}},\mathcal{D}} \right.} \right)}}{{p\left( {{{\mathbf{z}}_k}\left| {{{\mathbf{z}}_{1:k - 1}},\mathcal{D}} \right.} \right)}}
\label{Gau10}
\end{align}

The joint distribution of state prediction and measurement prediction is Gaussian \cite{wang2012gaussian}, i.e.,
\begin{align}
&p\left( {{{\mathbf{x}}_k},{{\mathbf{z}}_k}\left| {{{\mathbf{z}}_{1:k - 1}},\mathcal{D}} \right.} \right) 
\notag \\ &= N\left[ {\left( {\begin{array}{*{20}{c}}
  {{{{\mathbf{\hat x}}}_{k|k - 1}}} \\ 
  {{{{\mathbf{\hat z}}}_{k|k - 1}}} 
\end{array}} \right),\left( {\begin{array}{*{20}{c}}
  {{{\mathbf{P}}_{k|k - 1}}}&{{\mathbf{P}}_{k|k - 1}^{xz}} \\ 
  {{{\left( {{\mathbf{P}}_{k|k - 1}^{xz}} \right)}^{\top}}}&{{\mathbf{P}}_{k|k - 1}^z} 
\end{array}} \right)} \right]
\label{Gau11}
\end{align}

Following the conclusions in \cite{wang2012gaussian}, we substitute Eqs. (\ref{assum_meas}) and (\ref{Gau11}) into Eq. (\ref{Gau10}) to obtain updates of the state and covariance as shown in Eqs. (\ref{gau_state_up}) and (\ref{gau_cov_up}). The required ${{{\mathbf{\hat z}}}}_{k|k - 1}$, ${\mathbf{P}}_{k|k - 1}^z$, and ${\mathbf{P}}_{k|k - 1}^{xz}$ are obtained by the following procedure.

The mean value of the Gaussian measurement prediction is calculated as
\begin{align}
{{{\mathbf{\hat z}}}_{k|k - 1}} =& {\text{E}}\left[ {{h_k}\left( {{{\mathbf{x}}_k}} \right) + \Delta _k^h\left| {{{\mathbf{z}}_{1:k - 1}}} \right.} \right] 
\notag \\
=& {\iiint {\left( {{h_k}\left( {{{\mathbf{x}}_k}} \right) + \Delta _k^h} \right)}P_k^2d\Delta _k^hd{{\mathbf{c}}_k}d{{\mathbf{x}}_k}} 
\label{Gau12}
\end{align}
with
\begin{align}
P_k^2 = p\left( {\Delta _k^h\left| {{{\mathbf{x}}_k},\mathcal{D}} \right.} \right)p\left( {{{\mathbf{x}}_k},{{\mathbf{c}}_k}\left| {{{\mathbf{z}}_{1:k - 1}},\mathcal{D}} \right.} \right)
\label{Gau13}
\end{align}
Substituting $p\left( {\Delta _k^h\left| {{{\mathbf{x}}_{k - 1}},\mathcal{D}} \right.} \right)=\mathcal{N}( {\Delta _k^h;\hat \Delta _k^h,{\mathbf{P}}_k^h})$ into Eq. (\ref{Gau13}), Eq. (\ref{Gau12}) is computed as Eq. (\ref{Gau_meas_pred}).

Considering the measurement residuals is ${\sigma^z _k} = {{\mathbf{z}}_k} - {{{\mathbf{\hat z}}}_{k|k - 1}}$, the measurement prediction covariance is calculated as
\begin{align}
{\mathbf{P}}_{k|k - 1}^z =& {\text{E}}\left[ {{\sigma^z _k}(\sigma^z _k)^{\top}\left| {{{\mathbf{z}}_{1:k - 1}}} \right.} \right] 
\notag \\
=& \int \int {\int {M_k^zP_k^2d\Delta _k^hd{{\mathbf{c}}_k}d} } {{\mathbf{x}}_k}
\label{Gau17}
\end{align}
with
\begin{align}
M_k^z = \left( {{{h}_k}\left( {{{\mathbf{x}}_k}} \right) + \Delta _k^h + {{\mathbf{v}}_k} - {{{\mathbf{\hat z}}}_{k|k - 1}}} \right){\left(  \cdot  \right)^{\top}}
\label{Gau18}
\end{align}

Substituting $\mathcal{N}( {{\Delta _k^h};\hat \Delta _k^h,{\bf P}_k^f} )$ into Eq. (\ref{Gau7}) while considering Eq. (\ref{Gau3}), then we have Eq. (\ref{Gau_cov_pred}).

The mutual covariance of the state prediction and the measurement prediction is calculated as
\begin{align}
{\mathbf{P}}_{k|k - 1}^{xz} =& {\text{E}}\left[ {{\varepsilon^x _k}(\sigma^z _k)^{\top}\left| {{{\mathbf{z}}_{1:k - 1}}} \right.},\mathcal{D} \right] 
\notag\\
=& {\iiint {{\varepsilon^x _k}(\sigma^z_k)^{\top}P_k^2d{{ h}_k^c}d{{\mathbf{c}}_k}d{{\mathbf{x}}_k}} } 
\label{Gau20}
\end{align}

Substituting $\mathcal{N}( {{\Delta _k^h};\hat \Delta _k^h,{\bf P}_k^h} )$ into Eq. (\ref{Gau20}) while considering Eq. (\ref{Gau3}), then we have Eq. (\ref{Gau_xz_cov}). 

\bibliographystyle{IEEEtran}
\bibliography{ref}

% \begin{IEEEbiography}[{\includegraphics[width=1in,height=1.25in,clip,keepaspectratio]{fig/x1.png}}]{Name} Personal profile.
% \end{IEEEbiography}

% \begin{IEEEbiography}[{\includegraphics[width=1in,height=1.25in,clip,keepaspectratio]{fig/x2.png}}]{Name} Personal profile.
% \end{IEEEbiography}
	
% \begin{IEEEbiography}[{\includegraphics[width=1in,height=1.25in,clip,keepaspectratio]{fig/x3.jpg}}] {Name} Personal profile.
% \end{IEEEbiography}

% \begin{IEEEbiography}[{\includegraphics[width=1in,height=1.25in,clip,keepaspectratio]{fig/x4.pdf}}]{Name} Personal profile.
% \end{IEEEbiography}

% \begin{IEEEbiography}[{\includegraphics[width=1in,height=1.25in,clip,keepaspectratio]{fig/x5.png}}]{Name} Personal profile.
% \end{IEEEbiography}

\end{document}